\documentclass[prd,email,twocolumn,showpacs,showkeys,preprintnumbers,amsmath,amssymb,nofootinbib]{revtex4-1}

\usepackage{amsmath}
\usepackage{latexsym}

\def\[{\left\lbrack}
\def\]{\right\rbrack}

\def\({\left(}
\def\){\right)}

\newcommand{\be}{\begin{equation}}
\newcommand{\ee}{\end{equation}}
\newcommand{\ea}{\end{eqnarray}}
\newcommand{\ba}{\begin{eqnarray}}

\begin{document}

\title{Gauge invariance and dual equivalence of Abelian and non-Abelian \\ actions via dual embedding formalism}

\author{E. M. C. Abreu$^a$}
\email{evertonabreu@ufrrj.br}
\author{J. Ananias Neto$^b$} 
\author{A. C. R. Mendes$^b$}
\email{albert@fisica.ufjf.br} 
\author{C. Neves$^c$}
\email{clifford@fat.uerj.br} 
\author{W. Oliveira$^b$}
\email{wilson@fisica.ufjf.br}

\affiliation{${}^{a}$Grupo de F\' isica Te\'orica e Matem\'atica F\' isica, Departamento de F\'{\i}sica, Universidade Federal Rural do Rio de Janeiro\\
BR 465-07, 23890-971, Serop\'edica, RJ, Brazil\\
${}^{b}$Departamento de F\'{\i}sica, ICE, Universidade Federal de Juiz de Fora,\\
36036-330, Juiz de Fora, MG, Brazil\\
${}^{c}$Departamento de Matem\'atica e Computa\c{c}\~ao, Universidade do Estado do Rio de Janeiro\\
Rodovia Presidente Dutra, km 298, 27537-000, Resende, RJ, Brazil\\
\today\\}


\begin{abstract}
\noindent The concept of gauge invariance can be considered one of the most subtle and useful concept in theoretical physics since it can permit the comprehension of difficult systems in physics with an arbitrary choice of a reference frame at every instant of time.  It is always desirable to have a bridge between gauge invariant and noninvariant theories.  Once established, this kind of mapping between first-class (gauge invariant) and second-class systems, in Dirac's formalism can be considered as a sort of duality.  In this paper we investigate this ``duality" obtaining a gauge invariant theory starting with a noninvariant one.  We analyzed both Abelian and non-Abelian theories and the procedure used is the recent dual (also called symplectic) embedding formalism.  We believe that this method is the most convenient one since it is not plagued by the ambiguity problems that torments BFFT and other iterative methods.  We demonstrated exactly that this ``dualization" method used here does not require any special modification to handle with non-Abelian systems, which is also a new result described in this paper.  To prove the gauge invariance we used just the Dirac constraint technique.  It is relevant to say here that, although this work is lengthy, the majority of the results presented here are new in the literature.  The results that are not new were reproduced within a new perspective.
\end{abstract}

\maketitle

\pagestyle{myheadings}
\markright{{\it Gauge invariance and dual equivalence of Abelian and non-Abelian actions via dual embedding formalism}}

\newpage

\section{Introduction}
\renewcommand{\theequation}{1.\arabic{equation}}
\setcounter{equation}{0}

This work deals with the transformation of second-class systems into gauge theories.  Although there is an extensive literature about this subject, we obtain some results here that are altogether new.  In a self-sustained and pedagogical way we analyze gauge invariance and dual embedding issues.

A map with these conversion features can be understood as a kind of duality since we will show precisely that the final action have only first-class constraints.  This duality connection characterizes both systems as representing the same physical properties. We will discuss these ideas carefully later.

As well known, after the procedure, having only first-class constraints. the final system is a gauge theory, which can be thought as one in which the dynamical variables are determined with relation to a reference frame whose choice is arbitrary at every instant of time.  The relevance of a gauge theory, in few words, is that the physically important variables are those that are independent of the local reference frame \cite{ht}.  Whenever a change in the arbitrary reference frame causes a transformation of the variables involved we have the so-called gauge transformation.  This physical variables are then well known as gauge invariant variables.  Such gauge theories and gauge transformations are the cornerstones of the construction of the standard model, to mention only one of its successful applications.  We will talk with more detail about gauge theories and quantization in a moment.

Since the Hamiltonian formulation is considered by many as the more fundamental formulation of a physical theory we use the Dirac brackets in order to show exactly the gauge invariance of the final actions obtained.

{\bf Second-class $\rightarrow$ first-class.} Gauge theories have played an important role in field theories since they are related with the fundamental physical interactions of Nature.
In a more general sense, those theories have gauge symmetries defined by some relations called, in Dirac's language, first-class
constraints  \cite{PD}. The quantization of these theories demands a special care because the presence of  gauge symmetries indicate that exist some superfluous degrees of freedom, which must be eliminated (before or after) with the implementation of
a valid quantization process.

The quantization of first-class systems was formulated both in Dirac's  \cite{PD} and path
integral  \cite{FADDEEV} point of view. Later on, the path integral analysis was extended by Batalin, Fradkin and Vilkovisky \cite{FRADKIN} in order to preserve
the BRST symmetry  \cite{BRST}.

On the other hand, the covariant quantization of second-class systems is, in general, a difficult task because the Poisson brackets are
replaced by Dirac brackets. At the quantum level, the variables become operators and the Dirac brackets become commutators. Due to this,
the canonical quantization process is contaminated with serious problems such as ordering operator problems  \cite{order} and anomalies  \cite{RR} in
the context of nonlinear constrained systems and chiral gauge theories, respectively. In view of this fact, it seems that it is more natural and
safe to work out the quantization of second-class systems without invoking Dirac brackets. Actually, it was the strategy followed
by many authors over the last decades. The noninvariant system has been embedded in an extended phase space in order to change
the second-class nature of constraints into first-class.

In this way, the entire machinery  \cite{BRST,BV} for quantizing first class systems can be used. To implement this concept, Faddeev  \cite{FS} suggests to enlarge the phase space with the introduction of new variables to linearize the system, which were named after, as the Wess-Zumino (WZ) variables. This idea has been embraced by many authors and some methods were proposed and some constraint conversion formalisms, based on the Dirac method  \cite{PD}, were
constructed. Among them, the Batalin-Fradkin-Fradikina-Tyutin (BFFT)  \cite{BT} and the iterative  \cite{IJMP} methods were strong enough to be successfully applied to a great number of important physical systems. Although these techniques share the same conceptual basis  \cite{FS} and follow the Dirac framework  \cite{PD}, these constraint conversion methods were implemented following different directions. Historically, both BFFT and
the iterative methods were introduced to deal with linear systems such as chiral gauge theories  \cite{IJMP,many3} in order to eliminate
the gauge anomaly that hampers the quantization process.

In spite of the great success achieved by these methods, they have an ambiguity
problem  \cite{BN}. This problem naturally arise when the second-class constraints are converted into first class ones with the introduction
of WZ variables. Due to this, the process of constraint conversion process may become a hard task, as shown in  \cite{BN}.


\noindent {\bf Duality.} Duality is a very useful concept in field theory and statistical mechanics since there are very few analytic tools available for studying non-perturbative properties of systems with many degrees of freedom.  Always studied as one of the first applications of duality basics, the electromagnetic duality is a pedagogical and interesting example for the interested reader \cite{dualidade}.

Recently, the so-called gauging iterative Noether dualization method  \cite{ainrw} has
been shown to thrive in establishing some dualities between models  \cite{iw,iw2,iw3}.
This method is hinges on the traditional concept of a local lifting of a global symmetry and may be
realized by an iterative embedding of Noether counterterms.  However, this
method provides a strong suggestion of duality since it has been shown to give
the expected result in the paradigmatic duality between the so-called self-dual model  \cite{tpn} and the Maxwell-Chern-Simons
theory in three dimensions duality.  This correspondence was first established by Deser and Jackiw  \cite{dj}
and using the parent action approach  \cite{suecos}.

\noindent {\bf The paper.} We have organized this paper as follows. In section 2, we review the dual embedding formalism in order to settle the
notation and familiarize the reader with the fundamentals of the formalism.
In section 3, we will begin to make the application of the ideas discussed before in some Abelian models. We initiate with the Proca model in order to set up the general ideas discussed in Section 2.  We will see that a different zero-mode can bring us an equivalent action different from the literature.
After that, we apply this
formalism in two important models in high energy physics. The first one, in section 4, is the nonlinear sigma model (NLSM) \cite{NLSM}, which is an important
theoretical laboratory to learn the basics about asymptotically free field theories, as dynamical mass generation, confinement,
and topological excitations, which is expected in the realistic world of four-dimensional non-Abelian gauge theories. The second, in section 5, is the
bosonized chiral Schwinger model (CSM), which has attracted much attention over the last decade, mainly in the context of string
theories \cite{STRING}, and also due to the huge progress in understanding the physical meaning of anomalies in quantum field theories
achieved through the intensively study of this model. Through this section, we will first compute the Dirac brackets among the phase space
fields and after that, the gauge invariant version of the model will be obtained. Later on, the gauge symmetry will be investigated from the Dirac point of view.
Section 6 is devoted to an application of the formalism to the non-Abelian Proca model.
In this section, we will show that this gauge-invariant formalism does not require modifications to deal with non-Abelian models as demanded by the BFFT method.  In section 7, we analyze the symmetries of the rotational fluid model with a new extra term, which introduce a dissipative force into the system.  The objective is to promote an approximation to reality where we always have dissipation.
Our concluding observations and final comments are given in Section 8.

\section{The  dual embedding formalism}\label{s2}
\renewcommand{\theequation}{2.\arabic{equation}}
\setcounter{equation}{0}

The formalism is developed based on the symplectic framework  \cite{FJ,BC}, that is a modern way to handle with constrained systems. The basic object behind this
formalism is the symplectic matrix, i. e., if this matrix is singular, the model presents a symmetry, if it is not singular, the Dirac brackets can be
obtained. In this way, we propose to render nonsingular symplectic matrix to a singular one. It will be carried out introducing arbitrary
functions that depend on the original and WZ variables into the first-order Lagrangian. To appreciate this point, a brief review of the symplectic formalism will
be furnished and, after that, general ideas of the symplectic gauge-invariant formalism will be presented. This formalism, differently from the BFFT
and other iterative constraint conversion methods, does not require special modifications into the procedures to convert Abelian or non-Abelian set
of second-class constraints into first-class ones.

This technique follows the Faddeev-Shatashivilli's
suggestion  \cite{FS} and is set up on a contemporary framework to handle constrained
models, namely, the symplectic formalism  \cite{FJ}.
In the following lines, as we said above, we will try to keep this paper self-sustained reviewing the main steps of the dual embedding formalism.  We will follow closely the ideas contained in
 \cite{amnot}.

\bigskip

Let us consider a general noninvariant mechanical model whose dynamics is governed by a Lagrangian
${\cal L}(a_i,\dot a_i,t)$, (with $i=1,2,\dots,N$), where $a_i$ and $\dot a_i$
are the space and velocity variables, respectively. Notice that this model does not result in a loss of
generality nor physical content. Following the symplectic method the zeroth-iterative
first-order Lagrangian one-form is written as
 \begin{equation}
\label{2000}
{\cal L}^{(0)}dt = A^{(0)}_\theta d\xi^{(0)\theta} - V^{(0)}(\xi)dt\,\,,
\end{equation}
and the symplectic variables are
\be
\xi^{(0)\theta} =  \left\{ \begin{array}{ll}
                               a_i, & \mbox{with $\theta=1,2,\dots,N $} \\
                               p_i, & \mbox{with $\theta=N + 1,N + 2,\dots,2N ,$}
                           \end{array}
                     \right.
\ee
where $A^{(0)}_\theta$ are the canonical momenta and $V^{(0)}$ is the symplectic potential. From the Euler-Lagrange equations of motion, the symplectic tensor is obtained as
\begin{eqnarray}
\label{2010}
f^{(0)}_{\theta\beta} = {\partial A^{(0)}_\beta\over \partial \xi^{(0)\theta}}
-{\partial A^{(0)}_\theta\over \partial \xi^{(0)\beta}}\,\,.
\end{eqnarray}
If the two-form
$$f \equiv \frac{1}{2}f_{\theta\beta}d\xi^\theta \wedge d\xi^\beta$$

\noindent is singular, the symplectic matrix (\ref{2010}) has a zero-mode $(\nu^{(0)})$ that generates a new constraint when contracted with the gradient of the symplectic potential,
\begin{equation}
\label{2020}
\Omega^{(0)} = \nu^{(0)\theta}\frac{\partial V^{(0)}}{\partial\xi^{(0)\theta}}\,\,.
\end{equation}
This constraint is introduced into the zeroth-iterative Lagrangian one-form equation (\ref{2000}) through a Lagrange multiplier $\eta$, generating the next one
\begin{eqnarray}
\label{2030}
{\cal L}^{(1)}dt &=& A^{(0)}_\theta d\xi^{(0)\theta} + d\eta\Omega^{(0)}- V^{(0)}(\xi)dt,\nonumber\\
&=& A^{(1)}_\gamma d\xi^{(1)\gamma} - V^{(1)}(\xi)dt,\end{eqnarray}
with $\gamma=1,2,\dots,(2N + 1)$ and
\begin{eqnarray}
\label{2040}
V^{(1)}&=&V^{(0)}|_{\Omega^{(0)}= 0},\nonumber\\
\xi^{(1)_\gamma} &=& (\xi^{(0)\theta},\eta),\\
A^{(1)}_\gamma &=&(A^{(0)}_\theta, \Omega^{(0)})\,\,.\nonumber
\end{eqnarray}
As a consequence, the first-iterative symplectic tensor is computed as
\begin{eqnarray}
\label{2050}
f^{(1)}_{\gamma\beta} = {\partial A^{(1)}_\beta\over \partial \xi^{(1)\gamma}}
-{\partial A^{(1)}_\gamma\over \partial \xi^{(1)\beta}} \,\,.
\end{eqnarray}
If this tensor is nonsingular, the iterative process stops and the Dirac brackets
 among the phase space variables are obtained from the inverse matrix
 $(f^{(1)}_{\gamma\beta})^{-1}$ and, consequently, the Hamiltonian equation of
 motion can be formulated and solved, as discussed in  \cite{gotay}. It is well known
 that a physical system can be described at least classically in terms of a symplectic
 manifold ${\cal M}$. From a physical point of view, ${\cal M}$ is the phase space of the system while
 a nondegenerate closed 2-form $f$ can be identified as being the Poisson bracket. The
 dynamics of the system is  determined just specifying a real-valued function (Hamiltonian)
$H$ on the phase space, {\it i.e.}, one of these real-valued function
solves the Hamiltonian equation, namely,
\be \label{2050a1}
\iota(X)f=dH, \ee
and the classical dynamical trajectories of the
system in the phase space are obtained. It is important to mention
that if $f$ is nondegenerate, the equation (\ref{2050a1}) has a very unique
solution. The nondegeneracy of $f$ means that the linear map
$\flat:TM\rightarrow T^*M$ defined by $\flat(X):=\flat(X)f$ is an
isomorphism.  Due to this, the equation (\ref{2050a1}) is solved uniquely
for any Hamiltonian $(X=\flat^{-1}(dH))$. On the other hand, the
tensor has a zero-mode and a new constraint arises, indicating
that the iterative process goes on until the symplectic matrix
becomes nonsingular or singular. If this matrix is nonsingular,
the Dirac brackets will be determined naturally. 

In \cite{gotay}, the
authors consider in detail the case when $f$ is degenerate. The main idea of this embedding formalism is to introduce extra fields into the model in order to obstruct the solutions of the Hamiltonian equations of motion.
We introduce two arbitrary functions that hinge on the original phase space and on the WZ variables, namely, $\Psi(a_i,p_i)$ and $G(a_i,p_i,\eta)$, into the first-order one-form Lagrangian as follows
\be
\label{2060a}
{\tilde{\cal L}}^{(0)}dt = A^{(0)}_\theta d\xi^{(0)\theta} + \Psi d\eta - {\tilde V}^{(0)}(\xi)dt,
\ee
with
\be
\label{2060b}
{\tilde V}^{(0)} = V^{(0)} + G(a_i,p_i,\eta),
\ee
where the arbitrary function $G(a_i,p_i,\eta)$ is expressed as an expansion in terms of the WZ field, given by
\begin{equation}
\label{2060}
G(a_i,p_i,\eta)=\sum_{n=1}^\infty{\cal G}^{(n)}(a_i,p_i,\eta),,
\end{equation}
where
$${\cal G}^{(n)}(a_i,p_i,\eta)\sim\eta^n \,,$$
and satisfies the following boundary condition
\begin{eqnarray}
\label{2070}
G(a_i,p_i,\eta=0) = 0.
\end{eqnarray}
The symplectic variables were extended to also encompass the WZ variable $\tilde\xi^{(0)\tilde\theta} = (\xi^{(0)\theta},\eta)$ (with ${\tilde\theta}=1,2,\dots,2N+1$) and the first-iterative symplectic potential becomes
\begin{equation}
\label{2075}
{\tilde V}^{(0)}(a_i,p_i,\eta) = V^{(0)}(a_i,p_i) + \sum_{n=1}^\infty{\cal G}^{(n)}(a_i,p_i,\eta).
\end{equation}
In this context, the new canonical momenta are
\be
{\tilde A}_{\tilde\theta}^{(0)} = \left\{\begin{array}{ll}
                                  A_{\theta}^{(0)}, & \mbox{with $\tilde\theta$ =1,2,\dots,2N}\\
                                  \Psi, & \mbox{with ${\tilde\theta}$= 2N + 1}
                                    \end{array}
                                  \right.
\ee
and the new symplectic tensor is given by
\begin{equation}
{\tilde f}_{\tilde\theta\tilde\beta}^{(0)} = \frac {\partial {\tilde A}_{\tilde\beta}^{(0)}}{\partial \tilde\xi^{(0)\tilde\theta}} - \frac {\partial {\tilde A}_{\tilde\theta}^{(0)}}{\partial \tilde\xi^{(0)\tilde\beta}},
\end{equation}
that is
\be
\label{2076b}
{\tilde f}_{\tilde\theta\tilde\beta}^{(0)} =
\begin{pmatrix}
 { f}_{\theta\beta}^{(0)} & { f}_{\theta\eta}^{(0)}
\cr { f}_{\eta\beta}^{(0)} & 0
\end{pmatrix}.
\ee

To sum up, basically, we have two steps: the first one is addressed to compute $\Psi(a_i,p_i)$ while the second one is dedicated to the calculation of $G(a_i,p_i,\eta)$. In order to begin with the first step, we impose that this new symplectic tensor (${\tilde f}^{(0)}$) has a zero-mode $\tilde\nu$, consequently, we obtain the following condition
\begin{equation}
\label{2076}
\tilde\nu^{(0)\tilde\theta}{\tilde f}^{(0)}_{\tilde\theta\tilde\beta} = 0\,\,.
\end{equation}
At this point, $f$ becomes degenerated and in consequence, we introduce an obstruction to solve the Hamiltonian equation of motion given by equation (\ref{2050a1}). Assuming that the zero-mode $\tilde\nu^{(0)\tilde\theta}$ is
\begin{equation}
\label{2076a}
\tilde\nu^{(0)}=
\begin{pmatrix}
\mu^\theta & 1
\end{pmatrix},
\end{equation}
and using the relation given in (\ref{2076}) together with (\ref{2076b}), we have a system of equations,
\be
\label{2076c}
\mu^\theta{ f}_{\theta\beta}^{(0)} + { f}_{\eta\beta}^{(0)} = 0,
\ee
where
\be
{ f}_{\eta\beta}^{(0)} =  \frac {\partial A_\beta^{(0)}}{\partial \eta} - \frac {\partial \Psi}{\partial \xi^{(0)\beta}}\,\,.
\ee
The matrix elements $\mu^\theta$ are chosen in order to disclose the desired gauge symmetry. Note that in this formalism the zero-mode $\tilde\nu^{(0)\tilde\theta}$ is the gauge symmetry generator. It is worth to mention that this feature is important because it opens up the possibility to disclose the desired hidden gauge symmetry from the noninvariant model.  From relation (\ref{2076}) some differential equations involving $\Psi(a_i,p_i)$ are obtained, i. e., the equation (\ref{2076c}), and after a straightforward computation, $\Psi(a_i,p_i)$ can be determined.

In order to compute $G(a_i,p_i,\eta)$ following the second step of the method, it is mandatory that no constraints arise from the contraction of the zero-mode $(\tilde\nu^{(0)\tilde\theta})$ with the gradient of the potential ${\tilde V}^{(0)}(a_i,p_i,\eta)$. This condition generates a general differential equation, which reads as
\begin{widetext}
\begin{eqnarray}
\label{2080}
\tilde\nu^{(0)\tilde\theta}\frac{\partial {\tilde V}^{(0)}(a_i,p_i,\eta)}{\partial{\tilde\xi}^{(0)\tilde\theta}}\,&=&\, 0,\\
\mu^\theta \frac{\partial {V}^{(0)}(a_i,p_i)}{\partial{\xi}^{(0)\theta}} + \mu^\theta \frac{\partial {\cal G}^{(1)}(a_i,p_i,\eta)}{\partial{\xi}^{(0)\theta}}
\,+\, \mu^\theta\frac{\partial {\cal G}^{(2)}(a_i,p_i,\eta)}{\partial{\xi}^{(0)\theta}} &+& \dots
\,+\,\frac{\partial {\cal G}^{(1)}(a_i,p_i,\eta)}{\partial\eta} + \frac{\partial {\cal G}^{(2)}(a_i,p_i,\eta)}{\partial\eta} + \dots = 0\;\;, \nonumber \\
\mbox{}
\end{eqnarray}
\end{widetext}
that allows us to compute all the correction terms ${\cal G}^{(n)}(a_i,p_i,\eta)$ as functions of $\eta$. Notice that this polynomial expansion in terms of $\eta$ is equal to zero.  Consequently, all the coefficients for each order in $\eta$ must be identically null.
Given this, each correction term as function of $\eta$ is determined. For a linear correction term, we have
\begin{equation}
\label{2090}
\mu^\theta\frac{\partial V^{(0)}(a_i,p_i)}{\partial\xi^{(0)\theta}} + \frac{\partial{\cal
 G}^{(1)}(a_i,p_i,\eta)}{\partial\eta} = 0\,\,.
\end{equation}
For a quadratic correction term, we have
\begin{equation}
\label{2095}
{\mu}^{\theta}\frac{\partial{\cal G}^{(1)}(a_i,p_i,\eta)}{\partial{\xi}^{(0)\theta}} + \frac{\partial{\cal G}^{(2)}(a_i,p_i,\eta)}{\partial\eta} = 0.
\end{equation}
From these equations, a recursive equation for $n\geq 2$ can be chosen as,
\begin{equation}
\label{2100}
{\mu}^{\theta}\frac{\partial {\cal G}^{(n - 1)}(a_i,p_i,\eta)}{\partial{\xi}^{(0)\theta}} + \frac{\partial{\cal
 G}^{(n)}(a_i,p_i,\eta)}{\partial\eta} = 0,
\end{equation}
which permits us to compute the remaining correction terms as functions of $\eta$. This iterative process is successively repeated until (\ref{2080}) becomes identically zero.  Consequently, the extra term $G(a_i,p_i,\eta)$ is obtained explicitly. Then, the gauge invariant Hamiltonian, identified as being the symplectic potential, is obtained from
\begin{equation}
\label{2110}
{\tilde{\cal  H}}(a_i,p_i,\eta) = V^{(0)}(a_i,p_i) + G(a_i,p_i,\eta),
\end{equation}
and the zero-mode ${\tilde\nu}^{(0)\tilde\theta}$ is identified as the generator of an infinitesimal gauge transformation, given by
\begin{equation}
\label{2120}
\delta{\tilde\xi}^{\tilde\theta} = \varepsilon{\tilde\nu}^{(0)\tilde\theta},
\end{equation}
where $\varepsilon$ is an infinitesimal parameter.

In the following sections, we will apply the symplectic gauge-invariant formalism in some second-class constrained Hamiltonian systems, Abelian and non-Abelian.  We will see that the new results of the following sections demonstrate exactly that the dual (symplectic) embedding method can be seen as a mapping between the original action the final one.


\section{The Abelian Proca model}
\renewcommand{\theequation}{3.\arabic{equation}}
\setcounter{equation}{0}

The analysis of the following model clarify the physics and gives a deeper insight into the general formalism described in the last section. To this end, let us start with a simple Abelian
case which is the Proca model whose dynamics is ruled by the Lagrangian density,

\begin{equation}
\label{Proca1}
{\cal L} = - \,\,\frac{1}{4}F_{\mu\nu}F^{\mu\nu} + \frac{1}{2}\,m^2\,A^{\mu}A_{\mu},
\end{equation}
where $m$ is the mass of the $A_{\mu}$ field, $g_{\mu\nu} = diag(+---)$ and $F_{\mu\nu} = \partial_{\mu}A_{\nu} - \partial_{\nu}A_{\mu}$.
Observe that, as well known, the mass term breaks the gauge invariance of the usual Maxwell's theory. Hence, the Lagrangian density above represents a
second-class system.

To perform the dual embedding formalism the Lagrangian density is reduced to its first-order form as

\begin{equation}
\label{Proca3}
{\cal L}^{(0)} = \pi^{i}\dot{A_{i}} - V^{(0)},
\end{equation}
where the symplectic potential is

\be
\label{Proca3a}
V^{(0)} = \frac{1}{2}{\pi_{i}}^2 + \frac 14 F_{ij}^2 + \frac{1}{2}\,m^2\,{A_{i}}^2 - A_{0}(\partial_i\pi^{i} + \frac{1}{2}\,m^2\,A_{0}) ,
\ee
with 
$$\pi_i=\dot A_i - \partial_iA_0\,\,,$$ 

\noindent where $\partial_i=\frac{\partial}{\partial x^i}$ and the dot denote space and time derivatives,
respectively. The symplectic coordinates are $\xi_\alpha^{(0)}=(A_i,\pi_i,A_0)$ with the corresponding one-form canonical momenta given by

\begin{eqnarray}
\label{Proca4}
a_{A_i}^{(0)} &=& \pi_i, \nonumber \\
a_{\pi^i}^{(0)} &=& a_{A_0}^{(0)} = 0.
\end{eqnarray}
The zeroth-iteration symplectic matrix is

\begin{equation}
f^{(0)} = \left(
\begin{array}{ccc}
0           & -\delta_{ij} & 0 \\
\delta_{ji}&         0     & 0 \\
0           &         0     & 0
\end{array}
\right)\,\delta^{(3)}({\vec x} - {\vec y}),
\end{equation}
which is a singular matrix. It has a zero-mode that generates the following constraint,

\begin{equation}
\label{Proca5}
\Omega = \partial_i\pi^i + m^2A_0,
\end{equation}
identified as being the Gauss law.  We will include this constraint into the canonical part of the first-order Lagrangian ${\cal L}^{(0)}$
in (\ref{Proca3}) introducing a Lagrangian multiplier ($\beta$).  The first-iterated Lagrangian can be written in terms of $\xi_\alpha^{(1)} = (A_i,\pi_i,A_0, \beta)$ as

\begin{equation}
\label{Proca6}
{\cal L}^{(1)} = \pi^{i}\dot{A_{i}} + \Omega\dot{\beta} - V^{(1)},
\end{equation}
with the following symplectic potential,

\be
\label{Proca6a}
V^{(1)} = \frac{1}{2}{\pi_{i}}^2 + \frac 14 F_{ij}^2 + \frac{1}{2}\,m^2\,\({A_{0}}^2 + {A_{i}}^2\) - A_0\Omega.
\ee
The first-iterated symplectic matrix, computed as

\begin{equation}
f^{(1)}=\left(
\begin{array}{cccc}
0           & -\delta_{ij}  &  0   &   0 \\
\delta_{ji} &         0     &   0    &    \partial^y_i \\
0         &         0     &   0    &     m^2   \\
0       &        -\partial^x_j    &  -m^2     &     0
\end{array}
\right)\,\delta^{(3)}({\vec x} - {\vec y}),
\end{equation}
is a nonsingular matrix.  Consequently, the Proca model is not a gauge invariant field theory. As settle by the method, the Dirac brackets among the phase space fields are obtained from the inverse of the symplectic matrix, namely,

\ba
\label{dirac01}
\lbrace A_i(\vec x),A_j(\vec y)\rbrace^* &=& 0,\nonumber\\
\lbrace A_i(\vec x),\pi_j(\vec y)\rbrace^* &=& \delta_{ij}\delta^{(3)}(\vec x - \vec y),\\
\lbrace \pi_i(\vec x),\pi_j(\vec y)\rbrace^* &=& 0,\nonumber\\
\ea
with the following Hamiltonian,

\ba
\label{hamil01}
{\cal H} = V^{(1)}|_{\Omega=0} &=& \frac 12 \pi_i^2 - \frac {1}{2 m^2}\pi_i\partial^i\partial_j\pi^j + \frac 14 F_{ij}^2 + \frac 12 m^2 A_i^2,\nonumber\\
&=& \frac 12 \pi_iM^i_j\pi^j + \frac 14 F_{ij}^2 + \frac 12 m^2 A_i^2,
\ea
where the phase space metric is

\be
\label{metric01}
M^i_j = g^i_j - \frac{\partial^i\partial_j}{m^2}\,\,,
\ee
which completes the noninvariant analysis.

At this point we are ready to carry out the symplectic gauge-invariant formulation of the Abelian Proca model in order to disclose the gauge symmetry present in the model. To this end, we will extend the symplectic gauge-invariant formalism \cite{ANO}, proposed by three of us in
order to unveil, at that time, the gauge symmetry present on the Skyrme model. The basic concept behind the extended symplectic gauge-invariant formalism lives on the extension of the original phase space with the introduction of two arbitrary functions, $\Psi$ and $G$, where both rely on both the original phase space variables and the WZ variable $(\theta)$. The former ($\Psi$) is introduced into the kinetic sector and, the later (G), within the potential sector of the first-order Lagrangian. The process starts with the computation of $\Psi$ and finishes with the calculation of $G$.

In order to reformulate the Proca model as a gauge invariant field theory, we will start with the first-order Lagrangian ${\cal L}^{(0)},$ given in Eq. (\ref{Proca3}), with the arbitrary terms, given by,

\begin{equation}
\label{Proca6b}
{\tilde{\cal L}}^{(0)} = \pi^{i}\dot{A_{i}} + \dot\theta\Psi - {\tilde V}^{(0)},
\end{equation}
with

\be
\label{Proca6c}
{\tilde V}^{(0)} =  \frac{1}{2}{\pi_{i}}^2 + \frac 14 F_{ij}^2  + \frac{1}{2}\,m^2\,{A_{i}}^2 - A_{0}(\partial_i\pi^{i} + \frac{1}{2}\,m^2\,A_{0})  + G,
\ee
where $\Psi\equiv\Psi(A_i,\pi_i,A_0,\theta)$ and $G\equiv G(A_i,\pi_i,A_0,\theta)$ are the arbitrary functions to be determined. Now, the symplectic fields are ${\tilde\xi}^{(0)}_\alpha=(A_i,\pi_i,A_0,\theta)$ while the symplectic matrix is

\be
\label{matrix00}
f^{(0)} =
\begin{pmatrix}
0 & - \delta_{ij} & 0 & \frac{\partial\Psi_y}{\partial A^x_i}
\cr \delta_{ji} &  0 & 0 & \frac{\partial\Psi_y}{\partial \pi^x_i}\cr 0 & 0 & 0 & \frac{\partial\Psi_y}{\partial A^x_0}
\cr - \frac{\partial\Psi_x}{\partial A^y_j} & - \frac{\partial\Psi_x}{\partial \pi^y_j} & - \frac{\partial\Psi_x}{\partial A^y_0} & f_{\theta_x\theta_y}
\end{pmatrix}
\delta^{(3)}(\vec x - \vec y),
\ee
with

\be
\label{matrix01}
f_{\theta_x\theta_y} = \frac{\partial \Psi_y}{\partial \theta_x} - \frac{\partial \Psi_x}{\partial \theta_y},
\ee
where $\theta_x \equiv \theta(x)$, $\theta_y \equiv \theta(y)$, $\Psi_x \equiv \Psi(x)$ and $\Psi_y \equiv \Psi(y)$.

In order to unveil the hidden $U(1)$ gauge symmetry inside the Proca model, the symplectic matrix above must be singular, then, $\Psi\equiv(A_i,\pi_i,\theta)$. As established by the symplectic gauge-invariant formalism, the corresponding zero-mode $\nu^{(0)}(\vec x)$, identified as being the generator of the symmetry, satisfies the following relation,

\be
\label{matrix02}
\int \,\, d^3y \,\,\nu^{(0)}_\alpha(\vec x)\,\,f_{\alpha\beta}(\vec x - \vec y)= 0,
\ee
producing a set of equations that allows to determine $\Psi$ explicitly. At this point, it is very important to notice that the extended symplectic gauge-invariant formalism opens up the possibility to disclose the gauge symmetry of the physical model.  The zero-mode does not generate a new constraint, however, it determines the arbitrary function $\Psi$ and consequently, obtain the gauge invariant reformulation of the model. We will scrutinize the gauge symmetry related to the following zero-mode,

\be
\label{matrix03}
\bar\nu^{(0)} =
\begin{pmatrix}
\partial_i & 0 & 0 & 1
\end{pmatrix}.
\ee
Since this zero-mode and the symplectic matrix (\ref{matrix00}) must satisfy the gauge symmetry condition given in Eq. (\ref{matrix02}), a set of equations is obtained and after an integration, $\Psi$ is computed as

\be
\label{matrix04}
\Psi = - \partial_i\pi^i.
\ee
Hence, the symplectic matrix becomes

\be
\label{matrix05}
f^{(0)} =
\begin{pmatrix}
0 & - \delta_{ij} & 0 & 0
\cr \delta_{ji} &  0 & 0 & - \partial_i^y \cr 0 & 0 & 0 & 0
\cr 0 & \partial_j^x & 0 & 0
\end{pmatrix}
\delta^{(3)}(\vec x -\vec y),
\ee
which is singular by construction. Due to this, the first-order Lagrangian is

\begin{equation}
\label{Proca6ba}
{\tilde{\cal L}}^{(0)} = \pi^{i}\dot{A_{i}} - \partial_i\pi^i\dot\theta - {\tilde V}^{(0)},
\end{equation}
with ${\tilde V}^{(0)}$ given in Eq. (\ref{Proca6c}).

Now, we start with the second step of the formalism to transform the model into a gauge theory. The zero-mode $\bar\nu^{(0)}$ does not produce a
constraint when contracted with the gradient of the symplectic potential, namely,

\be
\label{matrix06}
\nu^{(0)}_\alpha\frac{\partial {\tilde V}^{(0)}}{\partial {\tilde\xi}_\alpha} = 0,
\ee
on the contrary, it produces a general equation that allows us to compute the correction terms in $\theta$ enclosed into $G(A_i,\pi_i,A_0,\theta)$,
given in Eq. (\ref{2080}). To compute the correction term linear in $\theta$, namely, ${\cal G}^{(1)}$, we pick up the following terms from
the general relation (\ref{2080}) given by,
\begin{widetext}
\be
\label{matrix07}
\int_x \,\[\partial^w_l\(m^2A^l(x)\delta^{(3)}(\vec x - \vec w) + \frac 12 F_{ij}(x)\frac{\partial F^{ij}(x)}{\partial A^l(w)}\) +
\frac{\partial{\cal G}^{(1)}(x)}{\partial\theta(w)}\] = 0.
\ee
\end{widetext}
After a straightforward calculation, the correction term linear in $\theta$ is,
\be
\label{matrix08}
{\cal G}^{(1)}  = - m^2\partial^iA_i \theta.
\ee
Substituting this result into the symplectic potential (\ref{Proca6c}), we have that

\ba
\label{matrix08a}
{\tilde V}^{(0)} &=&  \frac{1}{2}{\pi_{i}}^2 + \frac 14 F_{ij}^2 + \frac{1}{2}\,m^2\,{A_{i}}^2 - A_{0}(\partial_i\pi^{i} +
\frac{1}{2}\,m^2\,A_{0}) \nonumber \\
&-& m^2\partial^iA_i\theta .
\ea
However, the invariant formulation of the Proca model was not obtained yet because the contraction of the zero-mode (\ref{matrix03}) with
the symplectic potential above does not generate a null value. Due to this, higher order correction terms in $\theta$ must be computed. For
the quadratic term, we have,

\be
\label{matrix09}
\int_x\,\,\[ \partial^w_l\(- m^2\theta(x)\partial_x^l\delta^{(3)}(\vec x - \vec w)\)  +
\frac{\partial{\cal G}^{(2)}(x)}{\partial\theta(w)}\]= 0,
\ee
and after a direct calculation, we can write that,
\be
\label{matrix10}
{\cal G}^{(2)}  = +\, \frac 12 \,m^2\,(\partial_i\theta)^2\,\,.
\ee
Then, the first-order Lagrangian becomes,

\be
\label{matrix10a}
{\tilde{\cal L}} = \pi^i\dot A_i + \dot\theta\Psi - {\tilde V}^{(0)},
\ee
where the symplectic potential is

\ba
\label{matrix11}
{\tilde V}^{(0)} &=& \frac{1}{2}{\pi_{i}}^2 + \frac 14 F_{ij}^2 + \frac{1}{2}\,m^2\,{A_{i}}^2 - A_{0}(\partial_i\pi^{i} +
\frac{1}{2}\,m^2\,A_{0}) \nonumber \\
&-& m^2\partial^iA_i\theta + \frac 12 m^2\,\, \,(\partial_i\theta)^2.
\ea

Since the contraction of the zero-mode $({\bar\nu}^{(0)})$ with the symplectic potential above does not produce a new constraint,
a hidden symmetry is revealed.

To complete the gauge invariant reformulation of the Abelian Proca model, the infinitesimal gauge transformation will be computed also.
In agreement with the symplectic formalism, the zero-mode $\bar \nu^{(0)}$ is the generator of the infinitesimal gauge transformation
$(\delta{\cal O}=\varepsilon\bar\nu^{(0)})$, given by,

\begin{eqnarray}
\label{Proca16}
\delta A_i &=& - \partial_i\varepsilon,\nonumber\\
\delta \pi_i &=& 0,\\
\delta A_0 &=& 0,\nonumber\\
\delta\theta &=& \varepsilon,\nonumber
\end{eqnarray}
where $\varepsilon$ is an infinitesimal time-dependent parameter. Indeed, for the above transformations the invariant Hamiltonian,
identified as being the symplectic potential ${\tilde V}^{(0)}$, changes as

\be
\label{matrix12}
\delta{\cal H} = 0.
\ee

Now we will investigate the result from the Dirac point of view. The chains of primary constraints computed
from the Lagrangian (\ref{matrix10a}) are

\ba
\label{matrix13}
\phi_1 &=& \pi_0,\nonumber\\
\chi_1 &=& \partial_i\pi^i + \pi_\theta .
\ea
Next, these constraints will be introduced into the invariant Hamiltonian (\ref{matrix11}) through the Lagrange multipliers and so it can be
rewritten as

\be
\label{matrix14}
{\tilde V}^{(0)}_{primary} = {\tilde V}^{(0)} + \lambda_1\phi_1 + \gamma_1\chi_1.
\ee
The time stability condition for the primary constraint $\phi_1$ requires a secondary constraint, such as

\be
\label{matrix15}
\phi_2 = \partial_i\pi^i + m^2A_0,
\ee
and no more constraint appears from the time evolution of $\phi_2$. Now, the total Hamiltonian is written as

\be
\label{matrix16}
{\tilde V}^{(0)}_{total} = {\tilde V}^{(0)} + \lambda_1\phi_1 + \lambda_2\phi_2 + \gamma_1\chi_1.
\ee
Since the time evolution of $\phi_1$ just allows us to obtain the Lagrange multiplier $\lambda_2$, and the constraint $\chi_1$ has no time evolution $\dot\chi_1=0$, no more constraints arise. Hence, the gauge invariant model has three constraints $(\phi_1,\phi_2,\chi_1)$. The nonvanishing Poisson brackets among these constraints are,

\be
\lbrace \phi_1(x),\phi_2(y)\rbrace = - m^2\delta^{(3)}(\vec x - \vec y).
\ee
The Dirac matrix given by

\be
\label{matrix17}
C =
\begin{pmatrix}
0 & -1  & 0
\cr 1 & 0 & 0
\cr 0 & 0 & 0
\end{pmatrix}
m^2\delta(\vec x - \vec y),
\ee
is singular, indicating that the model has indeed, a gauge symmetry. However, it also has some nonvanishing Poisson brackets among the constraints, suggesting that the model has both first and second-class constraints. It is easy to check that $\chi_1$ is a first class constraint and $\phi_1$ and $\phi_2$ are second-class constraints. In accordance with the Dirac method, the set of second-class constraint must be taken equal to zero in a strong way, generating then the primary Dirac brackets among the phase space fields, given by

\ba
\label{matrix19}
\lbrace A_i(\vec x), \pi_j(\vec y)\rbrace^* &=& \delta_{ij}\delta^{(3)}(\vec x - \vec y),\nonumber\\
\lbrace \theta(\vec x), \pi_\theta(\vec y)\rbrace^* &=& \delta^{(3)}(\vec x - \vec y).
\ea
The gauge invariant version of the Abelian Proca model is then governed by the following invariant Hamiltonian, 

\ba
\label{matrix20}
{\tilde{\cal H}} &=& \frac{1}{2}{\pi_{i}}M^i_j\pi^j + \frac 14 F_{ij}^2  + \frac{1}{2}\,m^2\,{A_{i}}^2 - m^2\partial^iA_i\theta  \nonumber \\
&+&
\frac 12 m^2 \,(\partial_i\theta)^2,
\ea
whose phase space metric is

\be
\label{matrix21}
M^i_j = g^i_j - \frac{\partial^i\partial_j}{m^2}\,\,,
\ee
that has a first class constraint, $\chi_1$, which generates the infinitesimal transformations given in (\ref{Proca16}).

\section{The $O(N)$ invariant nonlinear sigma model}
\renewcommand{\theequation}{4.\arabic{equation}}
\setcounter{equation}{0}

In this subsection, the hidden symmetry present in the $O(N)$ nonlinear sigma model will be disclosed enlarging the phase space with the
introduction of WZ field {\it via} dual gauge-invariant formalism. We first apply the symplectic method to the original second-class
model in order to show the second-class nature of the model, and also to obtain the usual Dirac's brackets among the phase space fields.
After that, we unveil the hidden gauge symmetry of the model which dwells on the original phase space.

The $O(N)$ nonlinear sigma model in two dimensions is a free field theory for the multiplet $\sigma_a\equiv
(\sigma_1,\sigma_2,\dots,\sigma_n)$ satisfying a nonlinear constraint $\sigma_a^2=1$. This model has its dynamics governed by the
Lagrangian density

\begin{equation}
{\cal L}=\frac{1}{2}\,\partial_\mu\sigma^a\partial^\mu\sigma_a
- \frac{1}{2}\,\lambda\,\bigl(\sigma^a\sigma_a - 1\bigr),
\label{3001}
\end{equation}
where $\mu=0,1$ and $``a"$ is an index related to the $O(N)$ symmetry group.

In order to implement the symplectic method, the original second order Lagrangian in the velocity, given in (\ref{3001}), is reduced into
a first-order form, given by,

\begin{equation}
\label{3002}
{\cal L}^{(0)} = \pi_a\dot{\sigma}^a - V^{0},
\end{equation}
with

\be
\label{3002a}
V^{(0)} = \frac{1}{2}\,\pi^2_a + \frac{1}{2}\,\lambda\,\bigl(\sigma^2_a - 1\bigr) - \frac{1}{2}\,{\sigma^\prime_a} ^2,
\ee
where prime represent spatial derivatives, respectively. The symplectic coordinates are
$\xi_\alpha^{(0)}=(\sigma_a,\pi_a,\lambda)$ and the index ${(0)}$ indicates the zeroth-iteration. The symplectic tensor given by
Eq. (\ref{2010}) is computed in this case as

\begin{equation}
\label{3003}
f^{(0)} = \left(
\begin{array}{ccc}
0           & -\delta_{ab} & 0 \\
\delta_{ba} &         0     & 0 \\
0           &         0     & 0
\end{array}
\right)\,\delta(x-y).
\end{equation}
This matrix is singular, thus, it has a zero-mode, 

\begin{equation}
\label{3004}
\nu^{(0)} = \left(
\begin{array}{ccc}
{\bf 0} \\
{\bf 0} \\
1
\end{array}
\right).
\end{equation}
Contracting this zero-mode with the gradient of the symplectic potential $V^{(0)}$, given in Eq. (\ref{3002a}), the following constraint is
obtained,

\be
\label{3005a}
\Omega_1 = \sigma^2_a- 1.
\ee
In agreement with the symplectic formalism, this constraint must be introduced into the canonical sector of the first-order Lagrangian
(\ref{3002}) through a Lagrange multiplier $\rho$ and then, we obtain the first-iteration Lagrangian as

\begin{equation}
\label{3005}
{\cal L}^{(1)} = \pi_a\dot{\sigma}^a + \Omega_1\dot{\rho} - V^{(1)}\mid_{\Omega_1=0},
\ee
with

\be
\label{3006}
V^{1}\mid_{\Omega_1=0} =\frac{1}{2}\,\pi^2_a  + \frac{1}{2}\,{\sigma^\prime}^2_a.
\end{equation}
The symplectic coordinates are $\xi_\alpha^{(1)}=(\sigma_a,\pi_a,\rho)$ with the following one-form canonical momenta,

\begin{eqnarray}
\label{3007}
A_{\sigma_a}^{(1)} &=& \pi_a, \nonumber \\
A_{\pi_a}^{(1)} &=& 0,  \\
A_{\rho}^{(1)} &=& \bigl(\sigma^2_a - 1\bigr).\nonumber
\end{eqnarray}

The corresponding symplectic tensor $f^{(1)}$ given by,

\begin{equation}
\label{3008}
f^{(1)}=\left(
\begin{array}{ccc}
0           & -\delta_{ab} & \sigma_a \\
\delta_{ab} &         0     & 0 \\
-\sigma_b           &         0     & 0
\end{array}
\right)\,\delta(x-y),
\end{equation}
is singular, thus, it has a zero-mode that generates a new constraint, 

\be
\label{3009}
\Omega_2 = \sigma_a\pi^a.
\ee

Introducing the constraint $\Omega_2$ into the first-iteration Lagrangian (\ref{3005}) through a Lagrange multiplier $\zeta$, the
second-iteration Lagrangian is obtained as

\begin{equation}
\label{3010}
{\cal L}^{(2)} = \pi_a \dot{\sigma}^a + \dot{\rho}\bigl(\sigma^2_a - 1\bigr)+ \dot{\zeta}(\sigma_a\pi^a) - V^{(2)},
\end{equation}
with $V^{(2)}$ = $V^{(1)}\mid_{\Omega_1=0}$. The enlarged symplectic coordinates are $\xi_\alpha^{(2)}=(\sigma_a,\pi_a,\rho,\zeta)$ and the
new one-form canonical momenta are

\begin{eqnarray}
\label{formula22}
A_{\sigma_a}^{(2)} &=& \pi_a, \nonumber \\
A_{\pi_a}^{(2)} &=& 0, \nonumber \\
A_{\rho}^{(2)} &=& \sigma^2_a - 1,\nonumber \\
A_{\zeta}^{(2)} &=& \sigma_a \pi^a. \nonumber
\end{eqnarray}
The corresponding matrix $f^{(2)}$ is

\begin{equation}
\label{3011}
f^{(2)}=\left(
\begin{array}{cccc}
0           & -\delta_{ab}  &  \sigma_a   &   \pi_a \\
\delta_{ba} &         0     &   0    &    \sigma_a \\
-\sigma_b         &         0     &   0    &     0   \\
-\pi_b       &        -\sigma_b    &   0    &     0
\end{array}
\right)\,\delta(x-y),
\end{equation}
which is a nonsingular matrix. The inverse of $f^{(2)}$ furnish the usual Dirac brackets among the physical fields, given by,

\ba
\lbrace \sigma_a(x),\sigma_b(y)\rbrace^* &=& \lbrace \pi_a(x),\pi_b(y)\rbrace^* = 0, \nonumber\\
\lbrace \sigma_a(x),\pi_b(y)\rbrace^* &=&\(\delta_{ab} - \frac{\sigma_a\sigma_b}{\sigma^2}\)\delta(x - y),\\
\lbrace \pi_a(x),\pi_b(y)\rbrace^* &=& \frac{(\sigma_a\pi_b - \sigma_b\pi_a)}{\sigma^2}\delta(x - y).\nonumber
\ea
This means that the NLSM is not a gauge invariant theory.

At this stage we are ready to implement our proposal. In order to disclose the hidden symmetry present within the NLSM, the original phase
space will be extended with the introduction of WZ field following the symplectic gauge-invariant formalism. This
process is based on the introduction of two arbitrary functions, $\Psi(\sigma_a,\pi_a,\theta)$ and  $G(\sigma_a,\pi_a,\theta)$, into the
first-order Lagrangian as follows,

\be
\label{3012a}
{\tilde {\cal L}}^{(0)} = \pi_a\dot{\sigma}^a + \Psi\dot{\theta} - {\tilde V}^{(0)},
\ee
where the symplectic potential is

\be
\label{3013a}
{\tilde V}^{(0)} = \frac{1}{2}\,\pi^2_a + \frac{1}{2}\,\lambda\,\bigl(\sigma^2_a - 1\bigr) + \frac{1}{2}\,{\sigma^\prime_a} ^2 + G(\sigma_a,\pi_a,\theta),
\ee
with $G(\sigma_a,\pi_a,\theta)$ satisfying the relations given in Eqs.(\ref{2060}) and (\ref{2070}).

The symplectic coordinates are $\tilde \xi_\alpha^{(0)}=(\sigma_a,\pi_a,\lambda,\theta)$
with the following one-form canonical momenta,

\begin{eqnarray}
\label{3015}
\tilde A_{\sigma_a}^{(0)} &=& \pi_a, \nonumber \\
\tilde A_{\pi_a}^{(0)} &=& 0, \nonumber \\
\tilde A_{\lambda}^{(0)} &=& \frac 12 (\sigma^2_a - 1), \nonumber \\
\tilde A_{\theta}^{(0)} &=& 0.
\end{eqnarray}

As established by the symplectic gauge-invariant formalism, the corresponding matrix $\tilde f^{(0)}$, given by

\begin{equation}
\label{3016}
\tilde f^{(0)} =
\begin{pmatrix}
0  & -\delta_{ab}  &  0  &   \frac{\partial\Psi_y}{\partial\sigma^{x}_a}
\cr \delta_{ba} &  0 & 0  & \frac{\partial\Psi_y}{\partial\pi^{x}_a}
\cr 0 & 0 & 0  & \frac{\partial\Psi_y}{\partial\lambda^{x}}
\cr - \frac{\partial\Psi_x}{\partial\sigma^{y}_b} & - \frac{\partial\Psi_x}{\partial\pi^{y}_b}  & -\frac{\partial\Psi_x}{\partial\lambda^{y}} & f_{\theta_x\theta_y}
\end{pmatrix}
\delta(x-y),
\end{equation}

\noindent must be singular, this fixes the dependence relations of arbitrary function $\Psi$, namely,
$\frac{\partial\Psi_y}{\partial\lambda^{x}_a}=0$, {\it i.e}, $\Psi\equiv\Psi(\sigma_a,\pi_a,\theta)$.  
This matrix has a zero-mode, identified as being the gauge symmetry generator. To pull out the hidden symmetry, this zero-mode must satisfy the relation
(\ref{matrix02}), allowing then the computation of $\Psi$. 

Let us start considering the symmetry generated by the following zero-mode,

\begin{equation}
\label{3017}
\nu^{(0)}=\left(
\begin{array}{ccc}
{\bf 0} \\
\sigma_a \\
0 \\
1
\end{array}
\right).
\end{equation}
Since this zero-mode and the symplectic matrix (\ref{3016}) satisfy the relation (\ref{matrix02}), $\Psi$ is determined as

\be
\label{3017a}
\Psi = \sigma_a^2 + c,
\ee
where $``c"$ is a constant parameter. This completes the first step of our formalism.

The second step begins with the imposition that no more constraints are generated by the contraction of the zero-mode with the gradient of
the potential.  The correction terms as functions of $\theta$ can be explicitly computed. The first-order correction term in
$\theta$, ${\cal G}^{(1)}$, determined after an integration process, is

\begin{equation}
\label{3018}
{\cal G}^{(1)}(\sigma_a,\pi_a,\theta) = - \sigma_a\pi_a\theta.
\end{equation}
Substituting this expression into Eq. (\ref{3013a}), the new Lagrangian is

\begin{equation}
\label{3019}
\tilde {\cal L}^{(0)} = \pi_a\dot{\sigma}^a + \Psi\dot{\theta} - \frac{1}{2}\,{\sigma^\prime}^2_a - \frac{1}{2}\,\pi^2_a - \frac{1}{2}\,\lambda\,\bigl(\sigma^2_a - 1\bigr)  + \sigma_a\pi_a\theta.
\end{equation}

However, the model is not yet gauge invariant because the contraction of the zero-mode $\nu^{(0)}$ with the gradient of the potential
$V^{0}$ produces a non null value, indicating that it is necessary to compute the remaining correction terms ${\cal G}^{(n)}$ as functions of
$\theta$. It is carried out just demanding that the zero-mode does not generate a new constraint. It allows us to determine the second order
correction term ${\cal G}^{(2)}$, given by

\begin{eqnarray}
\label{3020}
{\cal G}^{(2)} = + \frac {1}{2} \sigma_a^2 \theta^2.
\end{eqnarray}
Substituting this result into the first-order Lagrangian (\ref{3019}), we have that,

\ba
\label{3021}
\tilde {\cal L}^{(0)} &=& \pi_a\dot{\sigma}^a + \Psi\dot{\theta} - \frac{1}{2}\,{\sigma^\prime} ^2_a -
\frac{1}{2}\,\lambda\,\bigl(\sigma^2_a - 1\bigr)  - \frac{1}{2}\,\pi^2_a  \nonumber \\
&+& \sigma_a\pi^a\theta - \frac {1}{2} \sigma_a^2 \theta^2.
\ea
Now the zero-mode $\nu^{(0)}$ does not produce a new constraint, consequently, the model has a symmetry and, in accordance with the
symplectic point of view, the generator of the symmetry is the zero-mode. Due to this, all correction terms ${\cal G}^{(n)}$ with
$n\geq 3$ are zero.

At this moment, we are interested in recovering the invariant second order Lagrangian from its first-order form given in Eq. (\ref{3021}).
To this end, the canonical momenta must be eliminated from the Lagrangian (\ref{3021}). From the equation of motion for $\pi_a$, the
canonical momenta are computed as

\begin{equation}
\label{3022}
\pi_a = \dot \sigma_a + \sigma_a\theta.
\end{equation}

\noindent Inserting this result into the first-order Lagrangian (\ref{3021}), we have the second order Lagrangian as

\begin{equation}
\label{3023}
\tilde {\cal L} = \frac{1}{2}\,\partial_\mu\sigma_a\partial^\mu\sigma^a - (\dot{\sigma_a}\sigma^a)\theta -
\frac{1}{2}\bigl(\sigma^2_a - 1\bigr)\lambda ,
\end{equation}
with the following gauge invariant Hamiltonian,

\begin{equation}
\label{3024}
\tilde {\cal H} = \frac{1}{2}\,\pi^2_a  + \frac{1}{2}\,{\sigma^\prime} ^2_a - (\sigma_a\pi^a)\theta +
\frac{1}{2}\lambda \bigl(\sigma^2_a - 1\bigr)+ \frac{1}{2}\,\sigma^2_a\theta^2 .
\end{equation}

\noindent Both Lagrangian (\ref{3023}) and Hamiltonian (\ref{3024}) are gauge invariant. From the Dirac point of view,
$\Omega_1$ arises as a secondary constraint from the temporal stability imposed on the primary constraints, $\pi_\lambda$ and
$\pi_\theta$,  and plays the role of the Gauss law, which generates the time independent gauge transformation. 

To proceed the quantization, we recognize the states of physical interest as those that are annihilated by $\Omega_1$. This gauge invariant formulation of the NLSM was also obtained by one of us in  \cite{WN} with the introduction of WZ fields, as established by the iterative method \cite{IJMP},
and by another authors using the BFFT formalism \cite{BGB}.

As established by the symplectic formalism, the zero-mode is identified as being the generator of the infinitesimal gauge transformations
$\delta\tilde \xi_\alpha^{(0)}=\varepsilon \nu^{(0)}$, namely,

\begin{eqnarray}
\label{3025}
\delta \sigma_a &=& 0,\nonumber\\
\delta \pi_a &=& \varepsilon \sigma_a ,\nonumber\\
\delta \lambda &=& 0,\\
\delta \theta &=& \varepsilon.\nonumber
\end{eqnarray}
For the transformation above the Hamiltonian changes as

\be
\label{3025a}
\delta{\cal H} = 0.
\ee

Similar results were also obtained in the literature using different methods based on Dirac's constraint
framework \cite{WN,BGB,JW1,JW2,HKP,NW}. However, these techniques are affected by some ambiguity problems, as said before, that naturally arise when the
second-class nature of the set of constraints turns into first-class with the introduction of the WZ fields. In our procedure, this
kind of problem does not arise and consequently the arbitrariness disappears.

Henceforth, we are interested in disclosing the hidden symmetry of the NLSM lying on the original phase space $(\sigma_a,\pi_a)$. To this
end, we use the Dirac method to obtain the set of constraints of the gauge invariant NLSM described by the Lagrangian (\ref{3023}) and
Hamiltonian (\ref{3024}), given by,

\begin{eqnarray}
\label{3026}
\phi_1 &=& \pi_\lambda,\nonumber\\
\phi_2 &=& - \frac 12(\sigma^2_a - 1),
\end{eqnarray}
and

\begin{eqnarray}
\label{3027}
\varphi_1 &=& \pi_\theta,\nonumber\\
\varphi_2 &=& \sigma_a\pi_a - \sigma_a^2\theta,
\end{eqnarray}

\noindent where $\pi_\lambda$ and $\pi_\theta$ are the canonical momenta conjugated to $\lambda$ and $\theta$, respectively. The corresponding Dirac
matrix is singular.  However, there are nonvanishing Poisson brackets among some constraints, indicating that there are both second-class
and first-class constraints.  This problem is solved separating the second-class constraints from the first-class ones through constraint
analysis. The set of first-class constraints is

\begin{eqnarray}
\label{3028}
\chi_1 &=& \pi_\lambda,\nonumber\\
\chi_2 &=& - \frac 12 (\sigma^2_a - 1) + \pi_\theta,
\end{eqnarray}
while the set of second-class constraints is

\begin{eqnarray}
\label{3029}
\varsigma_1 &=& \pi_\theta,\nonumber\\
\varsigma_2 &=& \sigma_a\pi_a - \sigma^2_a\theta .
\end{eqnarray}
Since the second-class constraints are assumed to be equal to zero in a strong way, and using the Maskawa-Nakajima theorem \cite{NM}, the Dirac
brackets are constructed as

\begin{eqnarray}
\label{3030}
\lbrace \sigma_i(x), \sigma_j(y)\rbrace^* &=& 0,\nonumber\\
\lbrace \sigma_i(x), \pi_j(y)\rbrace^* &=& \delta_{ij}\,\delta(x-y),\\
\lbrace \pi_i(x), \pi_j(y)\rbrace^* &=& 0.\nonumber
\end{eqnarray}
Hence, the gauge invariant Hamiltonian is rewritten as

\begin{eqnarray}
\label{3031}
\tilde {\cal H} &=& \frac{1}{2}\pi^2_a + \frac{1}{2}\,{\sigma^\prime} ^2_a
- \frac{1}{2}\frac{(\sigma_a\pi^a)^2}{\sigma_a\sigma^a} + \frac{\lambda}{2}(\sigma^2_a -1)\nonumber\\
&=& {1\over 2} \pi_iM_{ij}\pi_j + \frac{1}{2}\,{\sigma^\prime}^2_a + \frac{\lambda}{2}(\sigma_a^2 - 1),
\end{eqnarray}
where the phase space metric $M_{ij}$, given by

\be
\label{3032}
M_{ij} = \delta_{ij} - \frac {\sigma_i\sigma_j}{\sigma_k^2},
\ee
which is a singular matrix.   The set of first-class constraints becomes

\ba
\label{3033}
\chi_1 &=& \pi_\lambda,\nonumber\\
\chi_2 &=& - \frac 12(\sigma^2_a - 1).
\ea
Note that the constraint $\chi_2$, originally a second-class constraint, becomes the generator of gauge symmetries, satisfying the
first-class property

\begin{equation}
\label{3035}
\{\chi_2, \tilde{H} \} = 0.
\end{equation}
Due to this result, the infinitesimal gauge transformations are computed as

\ba
\label{3036}
\delta \sigma_a &=& \varepsilon\lbrace \sigma_a,\chi_2\rbrace = 0,\nonumber\\
\delta \pi_a &=& \varepsilon\lbrace\pi_a,\chi_2\rbrace=\varepsilon \sigma_a,\\
\delta\lambda &=& 0.\nonumber
\ea
where $\varepsilon$ is an infinitesimal parameter. It is easy to verify that the Hamiltonian (\ref{3031}) is invariant under these
transformations because $\sigma_a$ are eigenvectors of the phase space metric ($M_{ij}$) with null eigenvalues. In this section we
reproduce the results originally obtained in  \cite{KR} from an alternative point of view.

\section{The gauge invariant bosonized Chiral Schwinger Model}
\renewcommand{\theequation}{5.\arabic{equation}}
\setcounter{equation}{0}

It has been shown over the last decade that anomalous gauge theories in two dimensions can be consistently and unitarily quantized for
both Abelian \cite{RR,JR,many} and non-Abelian \cite{RR1,LR} cases. In this scenario, the two dimensional model that has been extensively
studied is the CSM. We start with the following Lagrangian density of the bosonized CSM with $a > 1$,

\ba
{\cal L} &=& -\frac 14 \, F_{\mu\nu}\,F^{\mu\nu} +
            \frac{1}{2}\,\partial_\mu\phi\,\partial^{\mu}\phi +
            q\,\left(g^{\mu\nu} -\,\epsilon^{\mu\nu}\right)\,
            \partial_{\mu}\phi\,A_{\nu} \nonumber \\
            &+&
           \frac{1}{2}\,q^2a\,A_{\mu}A^{\mu}\,.
\label{00000}
\ea

\noindent Here, $F_{\mu\nu} = \partial_{\mu}A_{\nu}-\partial_{\nu}A_{\mu}$, $g_{\mu\nu} =
\mbox{diag}(+1,-1)$ and $\epsilon^{01} = -\epsilon^{10} = \epsilon_{10} = 1$.   The symplectic method will be used here to quantize
the original second-class model.  Then, the Dirac brackets and the respective reduced Hamiltonian will be determined as well.
In order to implement the symplectic method, the original Lagrangian written in second order in velocity, given in (\ref{00000}), is reduced
into its first-order as follows,
\be
\label{mitra1}
L^{(0)} =\pi _\phi \dot{\phi} + \pi _1 \dot{A_1} - U^{(0)},
\end{equation}
where the zeroth-iterative symplectic potential $U^{(0)}$ is

\begin{eqnarray}
\label{mitra2}
U^{(0)}&=& {1\over 2}(\pi _1^2 +\pi _\phi ^2 +\phi ^{\prime 2}) - A_0\big[ \pi _1^\prime +
{1\over2}q^2(a-1)A_0  \nonumber \\
&+& q^2A_1 + q\pi _\phi + q\phi^\prime \big]\nonumber \\
&-& A_1 \big[-q\pi_\phi -{1\over 2}q^2(a+1)A_1 - q\phi^\prime \big],
\end{eqnarray}
where dot and prime represent temporal and spatial derivatives, respectively. The zeroth-iterative symplectic variables are
$\xi _\alpha^{(0)}=( \phi ,\pi _\phi,A_0 ,A_1 ,\pi _1)$ with
the following one-form canonical momenta $A_\alpha $,

\begin{eqnarray}
\label{00050}
A_\phi ^{(0)} &=& \pi _\phi, \nonumber \\
A_{A_1}^{(0)} &=& \pi_1, \\
A_{\pi _\phi}^{(0)} &=& A_{A_0}^{(0)}=A_{\pi _1}^{(0)} = 0.  \nonumber
\end{eqnarray}
the zeroth-iterative symplectic tensor is obtained as

\begin{equation}
\label{00060}
f^{(0)}(x,y)= \left( \begin{array}{ccccc}
0 & -1 & 0 & 0 & 0 \\
1 & 0 & 0 & 0 & 0 \\
0 & 0 & 0 & 0 & 0 \\
0 & 0 & 0 & 0 & -1 \\
0 & 0 & 0 & 1 & 0
\end{array} \right )\delta(x - y).
\end{equation}
This matrix is obviously singular, thus, it has a zero-mode that generates a constraint when contracted with the gradient of the
potential $U^{(0)}$, given by,

\begin{eqnarray}
\label{00080}
\Omega _1 &=& \nu_\alpha ^{(0)}{{\partial U^{(0)}}\over {\partial \xi _\alpha ^{(0)}}}\nonumber \\
&=&\pi _1^\prime +q^2(a-1)A_0 + q^2A_1 + q\pi _\phi + q\phi ^\prime,
\end{eqnarray}
that is identified as being the Gauss law, which satisfies the following Poisson algebra,

\begin{equation}
\label{00081}
\lbrace\Omega _1(x),\Omega _1(y)\rbrace = 0.
\end{equation}

\noindent Substituting the constraint $\Omega _1$ into the canonical sector of the
first-order Lagrangian through a Lagrange multiplier $\eta $, we have the first-iterative Lagrangian $L^{(1)}$, namely,

\begin{equation}
\label{00090}
L^{(1)} =\pi _\phi \dot{\phi} + \pi _1 \dot{A_1} + \Omega_1 \dot {\eta } - U^{(1)},
\end{equation}
with the first-order symplectic potential given by

\begin{eqnarray}
\label{00100}
U^{(1)}&=& {1\over 2}(\pi _1^2 +\pi _\phi ^2 +\phi ^{\prime 2}) - A_0 \big[ \pi _1^\prime +
{1\over2}q^2(a-1)A_0  \nonumber \\
&+& q^2A_1 + q\pi _\phi + q\phi^\prime \big]\nonumber \\
&-& A_1 \big[-q\pi_\phi -{1\over 2}q^2(a+1)A_1 - q\phi^\prime \big],
\end{eqnarray}
where $U^{(1)}=U^{(0)}$. Therefore, the symplectic variables become $\xi _\alpha^{(1)}=( \phi ,\pi _\phi, A_0 ,A_1 ,\pi _1, \eta )$
with the following one-form canonical momenta,

\begin{eqnarray}
\label{00110}
A_\phi ^{(1)} &=& \pi _\phi ,\nonumber \\
A_{A_1}^{(1)} &=& \pi _1 ,\nonumber \\
A_{A_0}^{(1)} &=& A_{\pi _\phi}^{(1)}= A_{\pi _1}^{(1)}= 0, \\
A_{\eta}^{(1)}&=& \pi _1^\prime +q^2(a-1)A_0 + q^2A_1 + q\pi _\phi + q\phi ^\prime. \nonumber
\end{eqnarray}
The corresponding matrix $f^{(1)}$ is then

\begin{widetext}
\begin{equation}
\label{00120}
f^{(1)}(x, y)= \left ( \begin{array}{cccccc}
0 & -1 & 0 & 0 & 0 & q\partial_y \\
1 & 0 & 0 & 0 & 0 & q \\
0 & 0 & 0 & 0 & 0 & q^2(a-1) \\
0 & 0 & 0 & 0 & -1 & q^2 \\
0 & 0 & 0 & 1 & 0 & \partial_y \\
-q\partial _x & - q & - q^2(a-1) & -q^2 & - \partial_x & 0
\end{array} \right )\delta (x-y),
\end{equation}
\end{widetext}

\noindent which is a nonsingular matrix. The inverse of $f^{(1)}(x,y)$ furnishes, after a straightforward calculation, the Dirac brackets among
the physical fields, 

\begin{eqnarray}
\label{00180}
\left\{\,\phi(x)\,,\,\phi(y)\,\right\}^{*}&=& 0, \nonumber \\
\left\{\,\phi(x)\,,\,\pi_\phi(y)\,\right\}^{*}&=& \delta (x - y)\, , \nonumber \\
\left\{\,\phi(x)\,,\, A_0(y)\,\right\}^{*}&=& -\frac{1}{q(a-1)}\,
                                       \delta (x - y)\, , \nonumber \\
\left\{\,\phi(x)\,,\, A_1(y)\,\right\}^{*}&=& 0, \nonumber \\
\left\{\,\phi(x)\,,\,\pi_1(y)\,\right\}^{*}&=& 0, \nonumber \\
\left\{\pi_\phi(x),\pi_\phi(y)\right\}^{*} &=& 0, \nonumber \\
\left\{\pi_\phi(x),A_0(y)\right\}^{*} &=& \frac{1}{q(a-1)}
                                       \partial_y \delta (x - y)\, , \nonumber \\
\left\{\pi_\phi(x),A_1(y)\right\}^{*} &=& 0, \nonumber \\
\left\{\pi_\phi(x),\pi_1(y)\right\}^{*} &=& 0, \nonumber \\
\left\{A_{1}(x),A_{0}(y)\right\}^{*} &=& -\frac{1}{q^{2}(a-1)}\,
                                        \partial_y\delta (x - y)\, , \\
\left\{A_{1}(x),A_{1}(y)\right\}^{*} &=& 0, \nonumber\\
\left\{A_{1}(x),\pi_{1}(y)\right\}^{*} &=& \delta (x - y)\, ,\nonumber \\
\left\{\pi_{1}(x),A_0(y)\right\}^{*} &=& \frac{1}{(a-1)}
                                           \delta (x - y)\,\,, \nonumber\\
\left\{\pi_{1}(x),\pi_{1}(y)\right\}^{*} &=& 0. \nonumber
\end{eqnarray}
This means that the model is not a gauge invariant theory.

The gauge symmetry inside the model will be disclosed via a new gauge-invariant formalism that does not require more than one
WZ field. The fundamental concept behind the symplectic gauge-invariant formalism dwells in the extension of the original phase space with
the introduction of two arbitrary function $\Psi(\phi,\pi_\phi,A_0,A_1,\pi_1,\theta)$ and $G(\phi,\pi_\phi,A_0,A_1,\pi_1,\theta)$,
depending on both the original phase space
variables and the WZ variable $\theta$, into the first-order Lagrangian, right on the kinetic and symplectic potential sector,
respectively. In this way, the first-order Lagrangian that
governs the dynamics of the bosonized CSM, given in Eq. (\ref{mitra1}), is rewritten as

\begin{equation}
\label{00300}
{\tilde L}^{(0)} =\pi _\phi \dot{\phi} + \pi _1 \dot{A_1} + \dot\theta\Psi - {\tilde U}^{(0)},
\end{equation}
where
\begin{widetext}
\begin{eqnarray}
\label{00310}
{\tilde U}^{(0)}&=& {1\over 2}(\pi _1^2 +\pi _\phi ^2 +\phi ^{\prime 2}) - A_0\big[ \pi _1^\prime +
{1\over2}q^2(a-1)A_0 + q^2A_1 + q\pi _\phi + q\phi^\prime \big]\nonumber \\
&-& A_1 \big[-q\pi_\phi -{1\over 2}q^2(a+1)A_1 - q\phi^\prime \big] + G(\phi , \pi _\phi, A_0, A_1, \pi _1, \theta ).
\end{eqnarray}
\end{widetext}

The gauge-invariant formulation encompasses two steps: one is dedicated to the computation of $\Psi$ while the other is addressed to the
calculation of $G$.

The enlarged symplectic variables are now ${\tilde \xi} _\alpha^{(0)}=( \phi ,\pi _\phi, A_0 ,A_1 ,\pi _1, \theta )$ with the following
one-form canonical momenta

\begin{eqnarray}
\label{00320}
{\tilde A}_\phi ^{(0)} &=& \pi _\phi ,\nonumber \\
{\tilde A}_{A_1}^{(0)} &=& \pi _1 ,\nonumber \\
{\tilde A}_{A_0}^{(0)} &=& {\tilde A}_{\pi _\phi}^{(0)}={\tilde A}_{\pi _1}^{(0)} = 0, \\
{\tilde A}_{\theta}^{(0)} &=& \Psi \nonumber.
\end{eqnarray}
The corresponding symplectic matrix ${\tilde f}^{(0)}$ reads

\begin{widetext}

\begin{equation}
\label{00325}
{\tilde f}^{(0)}=
\begin{pmatrix}
0 & -1 & 0 & 0 & 0 & \frac{\partial\Psi^y}{\partial \phi^x}
\cr 1 & 0 & 0 & 0 & 0 & \frac{\partial\Psi^y}{\partial \pi^x_\phi}
\cr 0 & 0 & 0 & 0 & 0 & \frac{\partial\Psi^y}{\partial A^x_0}
\cr 0 & 0 & 0 & 0 & -1 & \frac{\partial\Psi^y}{\partial A^x_1}
\cr 0 & 0 & 0 & 1 & 0 & \frac{\partial\Psi^y}{\partial \pi^x_1}
\cr - \frac{\partial\Psi^x}{\partial \phi^y} & - \frac{\partial\Psi^x}{\partial \pi^y_\phi}& -\frac{\partial\Psi^x}{\partial A^y_0} &
- \frac{\partial\Psi^x}{\partial A^y_1} & -
\frac{\partial\Psi^x}{\partial \pi^y_1}  & f_{\theta_x\theta_y}
\end{pmatrix}
\delta (x-y),
\end{equation}
\end{widetext}

where

\be
\label{00325a}
f_{\theta_x\theta_y} = \frac{\partial \Psi_y}{\partial \theta_x} - \frac{\partial \Psi_x}{\partial \theta_y},
\ee
with $\theta_x \equiv \theta(x)$, $\theta_y \equiv \theta(y)$, $\Psi_x \equiv \Psi(x)$ and $\Psi_y \equiv \Psi(y)$. Note that this matrix
is singular since $\frac{\partial\Psi^x}{\partial A^y_0} = 0$. Due to this, we conclude that
$\Psi\equiv\Psi(\phi,\pi_\phi,A_1,\pi_1,\theta)$.

To unveil the gauge symmetry hidden inside the model, we assume that this singular matrix has a zero-mode $(\nu^{(0)})$ that satisfies the
following relation,

\begin{equation}
\label{00326}
\int \nu^{(0)}_\alpha (x) {\tilde f}^{(0)}_{\alpha\beta}(x - y)\; d\,y = 0.
\end{equation}

\noindent From this relation a set of equations will be obtained and consequently, the arbitrary function $\Psi$ can be determined. 
We will now investigate the symmetry related to the following zero-mode,

\begin{equation}
\label{00327}
\bar\nu^{(0)} =
\begin{pmatrix}
q & - q\partial_x & 1 & \partial_x & - q^2  & - 1
\end{pmatrix},
\end{equation}
with bar representing a transpose matrix.

To start, we multiply the zero-mode (\ref{00327}) by the symplectic matrix (\ref{00325}), as shown in equation (\ref{00326}). Due to this,
some equations arise and after an integration $\Psi$ is determined as

\begin{equation}
\label{00327a}
\Psi = \pi^\prime + q\phi^\prime + q\pi_\phi + q^2A_1.
\end{equation}
The corresponding symplectic matrix (\ref{00325}) is rewritten as
\begin{equation}
\label{00330}
{\tilde f}^{(0)}= \left ( \begin{array}{cccccc}
0 & -1 & 0 & 0 & 0 &  q\partial_y\\
1 & 0 & 0 & 0 & 0 & q\\
0 & 0 & 0 & 0 & 0 & 0\\
0 & 0 & 0 & 0 & -1 & q^2 \\
0 & 0 & 0 & 1 & 0 & \partial _y\\
- q\partial_x & - q & 0 & - q^2 & - \partial_x & 0
\end{array} \right )\delta(x-y)
\end{equation}

\noindent which is obviously singular.   Consequently, it has a zero-mode that, by construction, is given by equation (\ref{00327}).

Now we start the second step of the method reformulating the  model as a gauge invariant theory. At this stage, the correction terms as functions of
$\theta$, embraced by the arbitrary function $G$, given in Eq. (\ref{2060}), will be computed. It is achieved just imposing that no more
constraints arise from the contraction of the zero-mode, given in Eq. (\ref{00327}), with the gradient of the symplectic potential,

\begin{equation}
\label{00331}
\nu^{(0)}_\alpha\frac{\partial {\tilde U}^{(0)}}{\partial \xi^{(0)}_\alpha} = 0.
\end{equation}

The first-order correction term in $\theta $, ${\cal G}^{(1)}$, is determined by,

\be
\label{00350}
{\cal G}^{(1)}(\phi , \pi_\phi, A_1, \pi_1, A_0, \theta ) = - \Omega_1 \theta + q^2(a-1)A^{\prime}_1\theta -
q^2\theta \pi _1,
\ee
after an integration. Substituting this expression into the equation (\ref{00300}), the new Lagrangian is obtained as

\begin{equation}
\label{00360}
{\tilde L}^{(0)} = \pi _\phi \dot{\phi} + \pi _1 \dot{A_1} + \Psi\dot {\theta } - {\tilde U}^{(0)},
\end{equation}
with
\begin{widetext}
\begin{eqnarray}
\label{00361}
{\tilde U}^{(0)}&=& {1\over 2}(\pi _1^2 +\pi _\phi ^2 +\phi ^{\prime 2}) - A_0\big[ \pi _1^\prime +{1\over2}q^2(a-1)A_0 + q^2A_1 + q\pi _\phi + q\phi^\prime \big]\nonumber \\
&-& A_1 \big[-q\pi_\phi -{1\over 2}q^2(a+1)A_1 - q\phi^\prime \big]
-\Omega_1 \theta + q^2(a-1)\theta^\prime A_1 - q^2\theta \pi _1.
\end{eqnarray}
\end{widetext}

The Lagrangian in (\ref{00360}) is not yet gauge invariant because the zero-mode $\bar\nu^{(0)}$ still generates new constraints, given by

\begin{equation}
\label{00370}
\nu_\alpha^{(1)}{{\partial {\tilde U}^{(0)}}\over {\partial \xi _\alpha ^{(0)}}} = q^2(a-1)\theta^{\prime\prime} -
q^2(a-1) \theta + q^4\theta,
\end{equation}

\noindent indicating that it is necessary to obtain the rest of the correction terms ${\cal G}^{(n)}$ as functions of $\theta $.  
This is accomplished just imposing that no more constraints are generated by the contraction of the zero-mode with the gradient of extended symplectic potential. 
It allows us to determine the second order correction term ${\cal G}^{(2)}$ as,

\begin{eqnarray}
\label{00380}
& &\nu_\alpha^{(0)}{{\partial {\tilde U}^{(0)}}\over {\partial \xi _\alpha ^{(0)}}} \nonumber \\
&=& - q^2(a-1)\theta + q^2(a-1)\theta^{\prime\prime} +
q^4\theta - {{\partial {\cal G}^{(2)}}\over {\partial \theta }}= 0, \nonumber \\
& &\Rightarrow\,{\cal G}^{(2)} = - {1\over 2}\;\; q^2(a-1){\theta ^{\prime}}^2 + {1\over 2}q^4{\theta }^2 - {1\over 2}q^2(a-1)\theta ^2. \nonumber \\
\mbox{}
\end{eqnarray}

Hence, the first-order Lagrangian (\ref{00360}) becomes

\begin{equation}
\label{00390}
{\tilde L}^{(0)} = \pi _\phi \dot{\phi} + \pi _1 \dot{A_1} + \Psi\dot\theta - {\tilde U}^{(0)},
\end{equation}
with the new symplectic potential

\begin{widetext}
\begin{eqnarray}
\label{00391}
{\tilde U}^{(0)} &=& {1\over 2}(\pi _1^2 +\pi _\phi ^2 +\phi ^{\prime 2}) - A_0\big[ \pi _1^\prime +{1\over2}q^2(a-1)A_0 + q^2A_1 +
q\pi _\phi + q\phi^\prime \big]\nonumber \\
&-& A_1 \big[-q\pi_\phi - {1\over 2}q^2(a+1)A_1 - q\phi^\prime \big] - \Omega_1\theta + q^2(a-1)\theta A^\prime_1 - q^2\theta \pi _1 \nonumber \\
&-&{1\over 2}\;\; q^2(a-1){\theta ^{\prime}}^2 + {1\over 2}q^4{\theta }^2 - {1\over 2}q^2(a-1)\theta ^2.
\end{eqnarray}
\end{widetext}

\noindent The contraction of the zero-mode $\bar\nu^{(0)}$ with the new symplectic potential above does not produce a new constraint. Consequently, the model has a symmetry and this zero-mode is the generator of the infinitesimal gauge transformations. Due to this, all correction
terms ${\cal G}^{(n)}$ with $n \geq 3$ are zero. The infinitesimal gauge transformations generated by the zero-mode $(\delta\xi_i=\varepsilon \nu^{(0)})$ are

\ba
\label{00395}
\delta \phi &=& q\varepsilon,\nonumber\\
\delta \pi_\phi &=&  q \varepsilon^{\prime},\nonumber\\
\delta A_0 &=& \varepsilon,\nonumber\\
\delta A_1 &=& - \varepsilon^{\prime},\\
\delta \pi_1 &=& - q^2\varepsilon,\nonumber\\
\delta \theta &=& - \varepsilon.\nonumber
\ea
It is easy to verify that the Hamiltonian, identified as being the new symplectic potential ${\tilde U}^{(0)}$, is invariant under these infinitesimal gauge transformation above, namely,

\begin{equation}
\label{00396}
\delta {\tilde U}^{(0)} = 0.
\ee

At this point, we are interested in the analysis of this result and we would like also to demonstrate that the anomaly was canceled. 
It will be carried out from Dirac's point of view. From the Lagrangian (\ref{00390}) the chains of primary constraints are computed as,

\begin{eqnarray}
\label{00406}
\varphi_1 &=& \pi_0,\nonumber\\
\chi_1 &=& - \pi_\theta  + \Psi.
\end{eqnarray}

\noindent These primary constraints will be introduced into the Hamiltonian through Lagrange multipliers. In this way, the primary
Hamiltonian, 

\be
\label{00407}
{\tilde U}^{(0)}_{primary} = {\tilde U}^{(0)} + \lambda_1\varphi_1 + \gamma_1\chi_1.
\ee
Since the constraint $\varphi_1$ has no time evolution, the following secondary constraint is required

\begin{equation}
\label{00407a1}
\varphi_2 =\Omega_1 - q^2(a-1)\theta,
\end{equation}
and no more constraints arise from the temporal stability condition. In this way, the total Hamiltonian is

\be
\label{00407a3}
{\tilde U}^{(0)}_{total} = {\tilde U}^{(0)} + \lambda_1\varphi_1 + \lambda_2\varphi_2 + \gamma_1\chi_1.
\ee

\noindent The time stability condition for the constraint $\chi_1$ just allows us to determine the Lagrange multiplier $\lambda_3$. In this way,
the gauge invariant version of the model has three constraints, namely, $\varphi_1$, $\varphi_2$ and $\chi_1$. The corresponding Dirac
matrix, given by,

\be
\label{00407a4}
C(x - y) = \left( \begin{array}{ccc}
0 & -q^2(a-1) & 0 \\
q^2(a-1) & 0 & q^2(a-1) \\
0 & -q^2(a-1) & 0 \\
\end{array}\right) \delta(x-y),
\ee
is singular. As the Dirac matrix is singular, the model has both first and second-class constraints. Through a constraint
combination, we obtain a set of first-class constraints such as,

\be
\label{00408a}
\tilde\chi_1 = - \pi_{\theta}  + \Psi - \pi_0 ,
\ee
and a set of second-class constraints, given by

\begin{eqnarray}
\label{00408}
\tilde\varphi_1 &=& \varphi_1,\nonumber\\
\tilde\varphi_2 &=& \Omega_1 - q^2(a-1)\theta .
\ea
It is again easy to verify that $\tilde\chi_1$ is a first-class constraint, identified as the Gauss law, while the others are second-class constraints. Note that the anomaly was removed. Hence, the Gauss law is also recognized as being the generator of the gauge transformation given in Eq. (\ref{00395}).

At this stage, we will compute the degrees of freedom of the gauge invariant model proposed by us. The model has one first-class and two second-class constraints and the phase space dimensions sum eight dependent fields, i.e.,
$(\phi,\pi_\phi, A_1,\pi_1,A_0,\pi_0,\theta,\pi_\theta)$. The first-class constraint eliminates two fields, while the second-class constraints eliminate two fields, summing then four fields eliminated.  Hence, the model has four independent fields, i.e., there are two independent degrees of freedom.

In order to obtain the Dirac brackets, the set of second-class constraints, $\tilde\varphi_1$ and $\tilde\varphi_2$ , will be assumed equal to zero in a strong way. After a straightforward computation, the Dirac brackets among the phase space fields are obtained as

\ba
\label{00408a1}
\lbrace\phi(x),\phi(y)\rbrace^* &=& 0,\nonumber\\
\lbrace\phi(x),\pi_\phi(y)\rbrace^* &=& \delta(x - y),\nonumber\\
\lbrace\phi(x),A_0(y)\rbrace^* &=& - \frac{1}{q(a-1)}\delta(x - y),\nonumber\\
\lbrace\phi(x),A_1(y)\rbrace^* &=& 0,\nonumber\\
\lbrace\phi(x),\pi_1(y)\rbrace^* &=& 0,\nonumber\\
\lbrace\phi(x),\theta(y)\rbrace^* &=& 0,\nonumber\\
\lbrace\phi(x),\pi_\theta(y)\rbrace^* &=& 0,\nonumber\\
\lbrace\pi_\phi(x),A_0(y)\rbrace^* &=& \frac{1}{q(a-1)}\partial_y\delta(x - y),\nonumber\\
\lbrace\pi_\phi(x),\pi_\phi(y)\rbrace^* &=& 0,\nonumber\\
\lbrace\pi_\phi(x),A_1(y)\rbrace^* &=& 0,\nonumber\\
\lbrace\pi_\phi(x),\pi_1(y)\rbrace^* &=& 0,\nonumber\\
\lbrace\pi_\phi(x),\theta(y)\rbrace^* &=& 0,\\
\lbrace\pi_\phi(x),\pi_\theta(y)\rbrace^* &=& 0,\nonumber\\
\lbrace A_1(x),A_0(y)\rbrace^* &=& -\frac{1}{q^2(a-1)}\partial_y\delta(x - y),\nonumber\\
\lbrace A_1(x),A_1(y)\rbrace^* &=& 0,\nonumber\\
\lbrace A_1(x),\pi_{1}(y)\rbrace^* &=& \delta(x - y),\nonumber\\
\lbrace A_1(x),\theta(y)\rbrace^* &=& 0,\nonumber\\
\lbrace A_1(x),\pi_\theta(y)\rbrace^* &=& 0,\nonumber\\
\lbrace\pi_1(x),A_0(y)\rbrace^* &=&  \frac{1}{(a-1)}\delta(x - y),\nonumber\\
\lbrace\pi_1(x),\pi_1(y)\rbrace^* &=& 0,\nonumber\\
\lbrace\pi_1(x),\theta(y)\rbrace^* &=& 0, \nonumber\\
\lbrace\pi_1(x),\pi_\theta(y)\rbrace^* &=& 0, \nonumber\\
\lbrace\theta(x),A_0(y)\rbrace^* &=& 0, \nonumber\\
\lbrace\theta(x),\pi_\theta(y)\rbrace^* &=& \delta(x - y), \nonumber\\
\lbrace\pi_\theta(x),A_0(y)\rbrace^*  &=& \delta(x - y),\nonumber\\
\lbrace\pi_\theta(x),\pi_\theta(y)\rbrace^* &=& 0. \nonumber
\ea

\noindent Note that the Dirac brackets among the original phase space fields were obtained before.
After this process, the model now have only one first-class constraint, 

\be
\label{00409a1}
\chi=\tilde\chi_1|_{\tilde\varphi_1 = \tilde\varphi_2 = 0} = - \pi_\theta  + \Psi
\ee
identified as the Gauss law, that satisfies the Poisson algebra, 

\be
\label{00409a2}
\lbrace\chi(x),\chi(y)\rbrace^* = 0.
\ee
In this way, the anomaly was eliminated, the symmetry is preserved, and the fundamental brackets among the original phase space fields were
reobtained. Note that the Gauss law is the generator of the gauge symmetry given in (\ref{00395}).

Once more, the number of the independent degrees of freedom matches with the result obtained in the second-class case. The invariant
model has a phase space $(\phi,\pi_\phi,A_1,\pi_1,\theta,\pi_\theta)$, totaling six dependent fields, and has a first-class constraint
which eliminates two fields.  Consequently, the model has two independent degrees of freedom.

At this point, we are interested in commenting about the consistency of the gauge invariant version of the bosonized CSM. To do this, the
remaining symmetry will be eliminated with the introduction of the unitary gauge-fixing term, given by,

\be
\label{00409a3}
\theta = 0.
\ee

\noindent Hence, both the noninvariant Hamiltonian and the corresponding Dirac brackets computed in the beginning of this section are reobtained, then recovering the anomaly. In this way, we conclude that the new symplectic gauge-invariant formalism does not change the physics contents present inside the model.

\section{Duality for the non-Abelian extension of the Proca model}
\renewcommand{\theequation}{6.\arabic{equation}}
\setcounter{equation}{0}

The non-Abelian extension of the Proca model has its dynamics governed by the following Lagrangian density,

\be
\label{N0000}
{\cal L} = -\frac 14 F_{\mu\nu}^aF_a^{\mu\nu} + \frac 12 A_\mu^a A_a^\mu,
\ee
with

\be
\label{N0010}
F_{\mu\nu}^a = \partial_\mu A_\nu^a - \partial_\nu A_\mu^a + g C^a_{bc}A_\mu^bA_\nu^c,
\ee
where the antisymmetric tensor $C^a_{bc} \: (C^a_{bc} = - C^a_{cb})$, are in fact a set of real constants, known as the structure constants of the
gauge group, and satisfy the following property, 

\be
\label{N0020}
C^a_{bc}C^d_{ae} + C^a_{eb}C^d_{ac} + C^a_{ce}C^d_{ab} = 0.
\ee

Since we are interested in analyzing the non-Abelian Proca model from the symplectic point of view, the Lagrangian will be reduced to its
first-order form as follows,

\ba
\label{N0030}
{\cal L}^{(0)} &=& \pi_a^i {\dot A}^a_i - \frac 12 (\pi_a^i)^2 + A_0^a\Omega_a - \frac 12 m^2 A_i^aA^i_a \nonumber \\
&-& \frac 12 m^2 A_0^aA^0_a - \frac 14 F_{kj}^aF_{kj}^a,
\ea
where

\be
\label{N0040}
\Omega_a = \partial_i\pi^i_a - gC^b_{ca} \pi_b^iA_i^c + m^2A_a^0.
\ee

\noindent The symplectic variables are given by

\ba
\label{N0050}
& &\xi_\alpha^a = (A_i^a,\pi_i^a,A_0^a),\nonumber\\
& &\mbox{} \nonumber \\
& &\!\!\!\!\!\!\!\!\!\!\!\!\!\!\!\!\!\!\!\!\!\!\!\!\!\!\!\!\!\!\!\!\!\!\!\!\!\!\!\!\!\mbox{and the symplectic matrix is} \nonumber\\
& &\mbox{} \nonumber \\
f^{(0)} &=&
\begin{pmatrix}
0 & - \delta_{ji}\delta^{ba} & 0
\cr \delta_{ij}\delta^{ab} & 0  & 0
\cr 0 & 0 & 0
\end{pmatrix}
\delta^{(3)}(\vec x - \vec y).
\ea

\noindent Since this matrix is singular, it has a zero-mode that generates the constraint $\Omega_a$, given by equation (\ref{N0040}). In agreement with
the symplectic method, this constraint is introduced into the kinetic sector of the first-order Lagrangian through a Lagrange multiplier,
namely,

\ba
\label{N0060}
{\cal L}^{(1)} &=& \pi_a^i {\dot A}^a_i + \Omega_a\dot\eta^a - \frac 12 (\pi_a^i)^2 + A_0^a\Omega_a - \frac 12 m^2 A_i^aA^i_a  \nonumber \\
&-&
\frac 12 m^2 A_0^aA^0_a - \frac 14 F_{kj}^aF_{kj}^a.
\ea
The new group of symplectic variables is $\xi_\alpha^a = (A_i^a,\pi_i^a,A_0^a,\eta^a)$, and the new symplectic matrix is

\begin{widetext}
\be
\label{N0070}
f^{(1)} =
\begin{pmatrix}
0 & - \delta_{ji}\delta^{ba} & 0 & - g C^{ab}_d\pi^d_i(y)\cr \delta_{ij}\delta^{ab} & 0  & 0 & \delta^{ab}\partial_i^y -
g C^{ab}_d A^d_i(y)
\cr 0 & 0 & 0 & m^2 \delta^{ab} \cr g C^{ba}_d\pi^d_j(x) & - \delta^{ba}\partial_j^x + g C^{ba}_d A^d_j(x) &
- m^2\delta^{ba} & 0
\end{pmatrix}
\delta^{(3)}(\vec x - \vec y).
\ee
\end{widetext}
This matrix is nonsingular and its inverse leads to the commutation relations among the dynamical variables, given by

\ba
\label{N0080}
\lbrace A_i^a(x), A_j^b(y)\rbrace &=& 0,\nonumber\\
\lbrace A_i^a(x), \pi_j^b(y)\rbrace &=& \delta^{ab}\delta_{ij}\delta^{(3)}(\vec x - \vec y),\nonumber\\
\lbrace \pi_i^a(x), \pi_j^b(y)\rbrace &=& 0,\\
\lbrace A_i^a(x), A_0^b(y)\rbrace &=& - \frac {1}{m^2} \delta^{ab}\partial_i^x \delta^{(3)}(\vec x - \vec y) \nonumber \\
&-&\frac {g}{m^2} C^{ab}_e A^e_i(x)\delta(\vec x - \vec y),\nonumber\\
\lbrace A_0^a(x), A_0^b(y)\rbrace &=& - 2 gm^2C^{ab}_eA^e_0(x)\delta^{(3)}(\vec x - \vec y),\nonumber\\
\lbrace \pi_i^a(x), A_0^b(y)\rbrace &=& - \frac {g}{m^2} C^{ab}_e \pi^e_i(x)\delta^{(3)}(\vec x - \vec y).\nonumber
\ea
This completes the analysis of the noninvariant description of the model.

The model now can be reformulated as a gauge invariant field theory. This will be carried out in the context of the symplectic
gauge-invariant formulation. In agreement with this formalism, the first-order Lagrangian (\ref{N0030}) can be rewritten as

\ba
\label{N0090}
{\tilde {\cal L}}^{(0)} &=& \pi_a^i {\dot A}^a_i + \Psi_a\dot\theta^a - \frac 12 (\pi_a^i)^2 + A_0^a\Omega_a  \nonumber \\
&-& \frac 12 m^2 A_i^aA^i_a - \frac 12 m^2 A_0^aA^0_a - \frac 14 F_{kj}^aF_{kj}^a - G, \nonumber \\
\mbox{}
\ea

\noindent where the arbitrary functions are

\ba
\label{N0100}
\Psi_a &\equiv& \Psi_a(A_i^a,\pi_i^a,A_0^a,\theta^a),\nonumber\\
G &\equiv& G (A_i^a,\pi_i^a,A_0^a,\theta^a)=\sum_{n=0}^\infty{\cal G}^{n}(A_i^a,\pi_i^a,A_0^a,\theta^a), \nonumber \\
\mbox{}
\ea

\noindent and where the $G$ function obeys a boundary condition given by,

\ba
\label{N0110}
G \equiv (A_i^a,\pi_i^a,A_0^a,\theta^a=0)={\cal G}^{0}(A_i^a,\pi_i^a,A_0^a,\theta^a=0)=0. \nonumber \\
\mbox{}
\ea
In this context, the corresponding symplectic matrix is

\be
\label{N0130}
f^{(0)} =
\begin{pmatrix}
0 & - \delta_{ji}\delta^{ba} & 0 & \frac{\partial\Psi_b(y)}{\partial A^a_i(x)}
\cr \delta_{ij}\delta^{ab} & 0  & 0 & \frac{\Psi_b(y)}{\partial \pi^a_i(x)}
\cr 0 & 0 & 0 & \frac{\partial\Psi_b(y)}{\partial A^a_0(x)}
\cr - \frac{\partial\Psi_a(x)}{\partial A^b_j(y)} & - \frac{\partial\Psi_a(x)}{\partial \pi^b_j(y)} & - \frac{\partial\Psi_a(x)}{\partial A^b_0(y)} & 0
\end{pmatrix}
\delta^{(3)}(\vec x - \vec y).
\ee

In order to determine the $\Psi_a$ functions, we analyze the symmetry related to the following zero-mode,

\be
\label{N0120}
\bar\nu^{(0)} =
\begin{pmatrix}
\partial^x_i & 0 & 0 & 1
\end{pmatrix}
\ee
with $\partial^x_i = \frac{\partial}{\partial x^i}$, which satisfies the following condition,

\be
\label{N0140}
\int_w\bar\nu^{(0)}(\vec x)f_{\alpha\beta}(\vec x - \vec w) = 0.
\ee
This condition produces a set of differential equations which allows us to compute the $\Psi_a$ function as

\be
\label{N0150}
\Psi_a =  - \partial_i \pi^i_a(x).
\ee
Consequently, the first-order Lagrangian is rewritten as

\be
\label{N0160}
{\tilde {\cal L}}^{(0)} = \pi_a^i {\dot A}^a_i - (\partial_i \pi^i_a)\dot\theta^a - {\tilde V}^{(0)},
\ee
where the symplectic potential is

\ba
\label{N170}
{\tilde V}^{(0)} &=& \frac 12 (\pi_a^i)^2 - A_0^a\Omega_a + \frac 12 m^2 A_i^aA^i_a + \frac 12 m^2 A_0^aA^0_a \nonumber \\
&+& \frac 14 F_{kj}^aF_{kj}^a + G.
\ea
It completes the first step of the symplectic gauge-invariant formulation.

To unveil the hidden symmetry inside the model, the zero-mode $\bar\nu^{(0)}$ does not generate new constraints, consequently, we have the following 
relation,

\be
\label{N0180}
\int_x \bar\nu^{(0)}_\alpha(w)\frac{\partial{\tilde V}(x)}{\partial \xi^a_\alpha(w)} = 0.
\ee

\noindent From this relation we can compute the whole set of correction terms as functions of $\theta^a$. 
The linear correction term in $\theta$ is computed as being, 

\be
\label{N0190}
\int_x \left\{\partial^w_i\frac{\partial V(x)^{(0)}}{\partial A_i^f(w)} + \frac{{\cal G}^{(1)}(x)}{\partial \theta^f(w)}\right\} = 0.
\ee
After an integration we have that

\ba
\label{N0200}
&&{\cal G}^{(1)}(x) \nonumber \\
&=&-\, g C_{fa}^b \partial^x_i(A_0^a(x)\pi_b^i(x))\theta^f(x) - m^2 (\partial^x_iA^i_f)\theta^f(x)\nonumber\\
 &-& \frac 12 \int_y \partial^y_i \left(F^a_{kj}(x)\frac{\partial F_a^{kj}(x)}{\partial A_i^f(y)}\right)\theta^f(y).
\ea
Now, we will compute the quadratic term, namely,

\be
\label{N0210}
\int_x \left\{\partial^w_i\frac{\partial {\cal G}^{(1)}(x)}{\partial A_i^f(w)} + \frac{{\cal G}^{(2)}(x)}{\partial \theta^f(w)}\right\} = 0.
\ee
Integrating this relation in $\theta^f(w)$, the quadratic correction term is obtained as

\ba
\label{N0220}
{\cal G}^{(2)}(x) &=&  \frac 12 m^2 (\partial_x^i \theta^f(x))^2 \nonumber \\
&+&
\frac 12\int_{\theta^f(x)}  \int_w \partial^w_i \int_y \left[(\partial^y_l{\cal A}^{il}_{fb})\theta^b(y)\right], \nonumber \\
\mbox{}
\ea
where

\ba
\label{N0230}
{\cal A}^{il}_{fb} = \frac{\partial F^a_{kj}(x)}{\partial A_i^f(w)} \frac{\partial F_a^{kj}(x)}{\partial A_l^b(y)} +
F^a_{kj}(x)\frac{\partial^2 F_a^{kj}(x)}{\partial A_i^f(w)\partial A_l^b(y)}. \nonumber \\
\mbox{}
\ea

In this way, two correction terms as functions of $\theta_a$ $({\cal G}^{(3)}(x)$ and ${\cal G}^{(4)}(x))$ remain to be computed. 
Let us compute the first one. It can be done from the following relation,

\ba
& &\int_z\left\{\partial_n^z \left[\frac 12 \int_{\theta^f(x)}\int_w \partial_k^w \int_y \partial^y_l\frac{\partial{\cal A}^{kl}_{fb}}{\partial A_n^g(z)} \theta^b(y)\right] \right. \nonumber \\
&+& \left. \frac{{\cal G}^{(3)}(x)}{\partial \theta_g(z)} \right\} = 0 \nonumber
\ea
and we can write that,
\ba
\label{N0240}
& &{\cal G}^{(3)}(x) \nonumber \\
&=& - \frac 12 \int_{\theta^g(z)} \int_z \partial_n^z \int_{\theta_f(x)} \int_w \partial_k^w \int_y \partial^y_l\frac{\partial{\cal A}_{fb}^{kl}}{\partial A^n_g(z)} \theta^b(y). \nonumber \\
\mbox{}
\ea
\noindent Finally, the last correction term is,

\begin{widetext}
\be
\label{N0250}
{\cal G}^{(4)}(x) = \frac 12 \int_{\theta^h(v)} \int_v \partial_i^v \int_{\theta_g(z)} \int_z \partial_n^z \int_{\theta_f(x)} \int_w
\partial_k^w \int_y \partial^y_l\frac{\partial^2{\cal A}_{fb}^{kl}}{\partial A^i_h(v)\partial A^n_g(z)} \theta^b(y).
\ee
\end{widetext}
Therefore, the gauge invariant first-order Lagrangian is

\be
\label{N0260}
{\tilde {\cal L}}^{(0)} = \pi_a^i {\dot A}^a_i - (\partial_i \pi^i_a)\dot\theta^a - {\tilde V}^{(0)},
\ee
where the gauge invariant Hamiltonian, identified as being the symplectic potential, is given by
\begin{widetext}
\ba
\label{N270}
{\tilde {\cal H}} &=& \frac 12 (\pi_a^i)^2 - A_0^a\Omega_a + \frac 12 m^2 A_i^aA^i_a + \frac 12 m^2 A_0^aA^0_a + \frac 14 F_{kj}^aF_{kj}^a
- g C_{fa}^b \partial^x_i(A_0^a(x)\pi_b^i(x))\theta^f(x)\nonumber\\ &-& m^2 (\partial^x_iA^i_f)\theta^f(x)
- \frac 12 \int_y \partial^y_i \left(F^a_{kj}(x)\frac{\partial F_a^{kj}(x)}{\partial A_i^f(y)}\right)\theta^f(y)
+ \frac 12 m^2 (\partial^i \theta^f(x))^2\nonumber\\
&+& \frac 12 \theta_f(w) \int_w \partial^w_i \int_y \left[(\partial^y_l{\cal A}_{fb}^{il})\theta^b(y)\right]
- \frac 12 \int_{\theta^g(z)} \int_z \partial_n^z \int_{\theta_f(x)} \int_w \partial_k^w \int_y \partial^y_l\frac{\partial{\cal A}_{fb}^{kl}}{\partial A^n_g(z)} \theta^b(y)\nonumber\\
&-& \frac 12 \int_{\theta^h(v)} \int_v \partial_i^v \int_{\theta_g(z)} \int_z \partial_n^z \int_{\theta_f(x)} \int_w
\partial_k^w \int_y \partial^y_l\frac{\partial^2{\cal A}_{fb}^{kl}}{\partial A^i_h(v)\partial A^n_g(z)} \theta^b(y).
\ea
\end{widetext}
This completes our proposal.

At this stage, we would like to disclose the hidden symmetry present inside the model using the Dirac point of view. To this end, we start with the set of primary constraints, 

\ba
\label{N0280}
\Omega_1^a &=& \partial^i\pi_i^a + \pi_\theta^a,\nonumber\\
\chi_1^a &=& \pi_0^a.
\ea
For the first set of constraints, the time stability condition is satisfied $(\dot\Omega_1^a =0)$.  For the second one, the following secondary constraints are required,

\be
\label{N0290}
\chi_2^a = \Omega^a - g C_f^{ba}\pi_b^i\partial_i\theta^f.
\ee
Due to this, the total Hamiltonian is

\be
\label{N0300}
{\cal H} = {\tilde{\cal H}} + \lambda_a^1\Omega^a_1 + \zeta_a^1\chi_1^a + \zeta_a^2\chi_2^a,
\ee
where $\lambda_a^1$, $\zeta_a^1$ and $\zeta_a^2$ are Lagrange multipliers. Since the Poisson brackets among those constraints are

\ba
\label{N0310}
\lbrace\Omega^a_1(x),\Omega^b_1(y)\rbrace &=& 0,\nonumber\\
\lbrace\Omega^a_1(x),\chi^b_1(y)\rbrace &=& 0,\nonumber\\
\lbrace\Omega^a_1(x),\chi^b_2(y)\rbrace &=& 0,\\
\lbrace\chi^a_1(x),\chi^b_2(y)\rbrace &=& -m^2\delta^{ab}\delta^{(3)}(\vec x - \vec y),\nonumber\\
\lbrace\chi^a_2(x),\chi^b_2(y)\rbrace &=& 2gC_{d}^{ab} \chi_2^d(x)\delta^{(3)}(\vec x - \vec y) \nonumber \\
&-& 2 gm^2C_{d}^{ab}A_0^d(x)\delta^{(3)}(\vec x - \vec y),\nonumber
\ea
no more constraints arise. Notice that some brackets above are null, indicating that there are both first and second-class constraints.
Indeed, the first-class constraint is $\Omega^a_1$ and the second-class are $\chi^a_1$ and $\chi^a_2$. In agreement with Dirac's
procedure, the second-class constraints can be taken equal to zero in a strong way, this allows us to compute the primary Dirac brackets.
Due to the Maskawa-Nakajima theorem \cite{NM}, the primary Dirac brackets among the phase space fields are canonical. To demonstrate this, the
brackets are computed explicitly. The Dirac matrix is

\be
\label{N0320}
C =
\begin{pmatrix}
0 & -m^2\delta^{cd} \cr m^2\delta^{dc} & B^{cd}
\end{pmatrix}
\delta^{(3)}(\vec x - \vec y),
\ee
with

\be
\label{N0330}
B^{cd} = 2gC_{b}^{cd} \chi_2^b(x)- 2 gm^2C_{b}^{cd}A_0^b(x).
\ee
The inverse of the Dirac matrix is

\be
\label{N0335}
C^{(-1)} = \frac {1}{m^2}
\begin{pmatrix}
\frac {B^{cd}}{m^2} & \delta^{cd} \cr - \delta^{dc} & 0
\end{pmatrix}
\delta^{(3)}(\vec x - \vec y).
\ee
In accordance with the Dirac process, the Dirac brackets among the phase space fields are obtained as

\ba
\label{N0340}
\lbrace A^a_i(x), A^b_j(y)\rbrace^* &=& 0, \nonumber\\
\lbrace A^a_i(x), \pi^b_j(y)\rbrace^* &=& \delta^{ab}\delta^{(3)}(\vec x - \vec y), \nonumber\\
\lbrace A^a_i(x), A^b_0(y)\rbrace^* &=& -\frac {1}{m^2}\partial^x_i\delta^{(3)}(\vec x - \vec y) \nonumber \\
&+& \frac {g}{m^2} C^{ab}_f A_i^f(x)\delta^{(3)}(\vec x - \vec y), \nonumber\\
\lbrace \pi^a_i(x), A^b_0(y)\rbrace^* &=& -\frac{1}{m^2} g C^{ab}_e\pi^e_i\delta^{(3)}(\vec x - \vec y),\nonumber\\
\lbrace \pi^a_i(x), \pi^b_j(y)\rbrace^* &=& 0, \nonumber\\
\lbrace A^a_0(x), A^b_0(y)\rbrace^* &=& -\frac{g}{m^2} C_e^{ab}A^e_0(x)\delta^{(3)}(\vec x - \vec y), \nonumber \\
\lbrace A^a_i(x), \theta^b(y)\rbrace^* &=& 0,\\
\lbrace \pi^a_i(x), \theta^b(y)\rbrace^* &=& 0,\nonumber\\
\lbrace A^a_0(x), \theta^b(y)\rbrace^* &=& 0,\nonumber\\
\lbrace A^a_i(x), \pi_\theta^b(y)\rbrace^* &=& 0,\nonumber\\
\lbrace \pi^a_i(x), \pi_\theta^b(y)\rbrace^* &=& 0,\nonumber\\
\lbrace A^a_0(x), \pi_\theta^b(y)\rbrace^* &=& \frac {g}{m^2} C_e^{ab}\partial^x_i\pi^e_(x)\delta^{(3)}(\vec x - \vec y),\nonumber\\
\lbrace \theta^a(x), \pi_\theta^b(y)\rbrace^* &=& \delta^{ab}\delta^{(3)}(\vec x - \vec y),\nonumber\\
\lbrace \theta^a(x), \theta^b(y)\rbrace^* &=& 0,\nonumber\\
\lbrace \pi_\theta^a(x), \pi_\theta^b(y)\rbrace^* &=& 0.\nonumber
\ea

\noindent Finally, the infinitesimal gauge transformations are obtained, namely,

\ba
\label{N0350}
\delta A_i^a &=& - \partial_i^x\varepsilon^a,\nonumber\\
\delta \pi_i^a &=& 0,\nonumber\\
\delta A_0^a &=& 0,\\
\delta \theta^a &=& \varepsilon^a,\nonumber\\
\delta \pi_\theta^a &=& 0,\nonumber
\ea
which lead to the invariant Hamiltonian.

To demonstrate that the gauge invariant formulation of the non-Abelian Proca model is dynamically equivalent to the original noninvariant
model, the symmetry is fixed by using the unitary gauge fixing procedure, 

\be
\label{N0360}
\varphi^a = \theta^a\approx 0,
\ee
which leads to the bracket below,

\be
\label{N0370}
\lbrace \Omega^a_1(x),\varphi^b(y)\rbrace = - \delta^{ab}\delta^{(3)}(\vec x - \vec y).
\ee
Due to this, a new Dirac brackets must be computed. The corresponding Dirac matrix for this set of constraints is

\be
\label{N0380}
C =
\begin{pmatrix}
0 & -1 \cr 1 & 0
\end{pmatrix}
\delta^{(3)}(\vec x - \vec y).
\ee
Using the inverse of this matrix, the Dirac brackets among the physical phase space fields are computed, which is equal to the one calculated in the original description given by equation (\ref{N0080}).
This result demonstrate that the symplectic formalism can be seen in fact as a dual mapping between the original theory and the final one.


\section{Hidden symmetry in the rotational fluid model}
\renewcommand{\theequation}{7.\arabic{equation}}
\setcounter{equation}{0}

Recently, some of us have proposed a Wess-Zumino (WZ) gauge-invariant version for the isentropic irrotational fluid model
 \cite{01}. In that work, we have demonstrated that the irrotational fluid model has a set of dynamically equivalent WZ gauge invariant versions. Further, in that paper the extra global symmetries, namely, Galileo antiboost and time rescaling, first obtained in \cite{02}, were lifted to local symmetries.

In this section, we propose an investigation of the symmetries of the rotational fluid model, but now with an extra term, like $k\rho\(\partial_i\theta+\alpha\partial_i\beta \)^2$, where $k$ is a constant, $\rho\equiv \rho(t,\vec r)$ is the mass density and $\theta\equiv \theta(t,\vec r)$ is the velocity potential. This term introduces a dissipative force into the model. The main motivation is that in the real world there is always dissipation, albeit sometimes extremely small. Our purpose is, to verify if the extra symmetries found in \cite{01} are broken with the introduction of this new term in the model. We will introduce a final result that is new in the literature,

As was demonstrated in \cite{01}, the isentropic irrotational fluid model has a dynamically equivalent family of WZ gauge-invariant descriptions. Thus, in order to establish our ideas, we are going to investigate how the inclusion of a dissipative term affects the dynamics of the fluid model considering only one of these zero modes presented in \cite{01}. Note that in the prescription of the symplectic formalism, for each zero-mode chosen we will have a gauge-invariant version for the model.  

In the next subsection, the scalar rotational fluid theory will be analyzed from the symplectic point of view \cite{FJ}, and the fundamental Dirac brackets of the fields will be computed. 

\subsection{Symplectic analysis of the rotational fluid}\label{s2.1}

In this section, we will analyze the rotational fluid dynamic model from the dual embedding  point of view. 

It is well known that systems that have vorticity and/or viscosity present Casimir invariants which obstruct the construction of a canonical formalism for fluid, as demonstrated in \cite{RJ}. However, this obstruction can be eliminated using the Clebsch parameters, as shown by Lin \cite{Lin} and by two of us in \cite{NW2}. In fact, with the introduction of the Clebsch parameters, it is possible to obtain a Lagrangian density for the rotational fluid with dissipation, in 3-dimensional, as being
\be
\label{002}
{\cal L} = -\rho(\dot \theta + \alpha\dot\beta) - V,
\ee
where the symplectic potential is
\be
\label{03}
V =\frac {1}{2}(1-k)\rho (\partial_i \theta+ \alpha\partial_i \beta) (\partial^i\theta+ \alpha\partial^i\beta) + V(\rho).
\ee
The symplectic coordinates are $\xi^{(0)}=(\rho, \theta, \alpha, \beta)$ with the corresponding zeroth-iterative one-form canonical momenta given by
\begin{eqnarray}
\label{04}
A_{\rho}^{(0)} &=& 0, \nonumber \\
A_{\theta}^{(0)} &=& -\rho,\\
A_{\alpha}^{(0)} &=& 0, \nonumber\\
A_{\beta}^{(0)} &=& - \alpha\rho. 
\end{eqnarray}

\noindent The zeroth-iteration symplectic matrix, given by
\begin{equation}
f^{(0)} = \left(
\begin{array}{cccc}
0           & -\delta(\vec r -\vec r^{\prime}) & 0 & -\alpha\\
\delta(\vec r -\vec r^{\prime})&         0     & 0 & 0\\
0 &         0      &0 & -\rho\\
\alpha & 0 & \rho & 0 
\end{array}
\right),
\end{equation}
is a nonsingular matrix and, consequently, the  model is not a gauge-invariant field theory. As settle by the symplectic formalism \cite{FJ}, the Dirac brackets of the phase space fields are acquired from the inverse of the symplectic matrix, and are given by 
\ba
\label{05}
\lbrace \rho(\vec r),\theta(\vec r^{\prime})\rbrace^* &=&\delta(\vec r -\vec r^{\prime}),\nonumber\\
\lbrace \theta(\vec r),\alpha(\vec r^{\prime})\rbrace^* &=& \frac{\alpha}{\rho}\delta(\vec r -\vec r^{\prime}),\\
\lbrace \alpha(\vec r),\beta(\vec r^{\prime})\rbrace^* &=& \frac{1}{\rho}\delta(\vec r -\vec r^{\prime}),\nonumber
\ea
while the remaining brackets are null. This completes the noninvariant analysis.

In the next subsection, we will apply the symplectic embedding formalism \cite{ANO} in the 3-dimensional fluid dynamical model with dissipation and, as a consequence, we will obtain a gauge-invariant version for this model. Although the symplectic formalism does not restrain the dimension of the model, we choose a 3-dimensional description for the rotational fluid in order to put our work in a correct perspective with others. 

\subsection{Obtaining the gauge-invariant version of the model}\label{s3}

In this section, we will obtain the WZ gauge-invariant version of the fluid theory.   We will consider initially a general interaction potential and after that, a specific potential $(V=g/\rho)$ will be chosen. Following the prescription of the formalism, two arbitrary functions, $\Psi$ and $G$, depending on the original phase space fields and the WZ field $(\eta)$, must be added to the model. The former is introduced into the kinetic sector and the latter one into the potential sector of the first-order Lagrangian. The process starts with the computation of $\Psi$ and finishes with the computation of $G$.

In order to reformulate the model as a gauge-invariant field theory, let us start with the first-order Lagrangian ${\cal L}^{(0)}$, equation (\ref{002}), with additional arbitrary terms $(\Psi, G)$, 
\begin{equation}
\label{06}
{\tilde{\cal L}}^{(0)} = -\rho(\dot \theta + \alpha\dot\beta) + \Psi\dot\eta - {\tilde V}^{(0)},
\end{equation}
where
\be
\label{07}
{\tilde V}^{(0)} =\frac{1}{2}(1-k)\rho (\partial_i \theta + \alpha\partial_i \beta) (\partial^i\theta + \alpha\partial^i\beta) + V(\rho) + G,
\ee
and where $\Psi\equiv\Psi(\rho,\theta)$ and $G\equiv G(\rho,\theta,\eta)$ are arbitrary functions to be determined. Now, the symplectic coordinates are ${\tilde\xi}^{(0)}=(\rho,\theta,\alpha,\beta,\eta)$ while the symplectic matrix is
\be
\label{08}
{\tilde f}^{(0)} = \left(
\begin{array}{ccccc}
 0 & - \delta(\vec r - \vec r^{\prime}) &  0 & -\alpha & {\frac{\delta\Psi_{\vec r^{\prime}}}{\delta \rho(\vec r)}}\\ \nonumber
\delta(\vec r - \vec r^{\prime}) &  0 & 0& 0 & {\frac{\delta\Psi_{\vec r^{\prime}}}{\delta \theta(\vec r)}} \\ 
0& 0 & 0 & -\rho & {\frac{\delta\Psi_{\vec r^{\prime}}}{\delta \alpha(\vec r)}}\\

\alpha & 0 & \rho  & 0 & {\frac{\delta\Psi_{\vec r^{\prime}}}{\delta \beta(\vec r)}}\\

 - {\frac{\delta\Psi_{\vec r}}{\delta \rho(\vec r^{\prime})}} & - {\frac{\delta\Psi_{\vec r}}{\delta \theta(\vec r^{\prime})}} &  - {\frac{\delta\Psi_{\vec r}}{\delta \alpha(\vec r^{\prime})}}& - {\frac{\delta\Psi_{\vec r}}{\delta \beta(\vec r^{\prime})}}&0\nonumber
\end{array}\right),
\ee
where $\Psi_{\vec r} \equiv \Psi(\rho(\vec r),\theta(\vec r))$ and $\Psi_{\vec r^{\prime}} \equiv \Psi(\rho(\vec r^{\prime}),\theta(\vec r^{\prime}))$.

As established by the dual embedding method, the corresponding zero-mode $\tilde\nu^{(0)}(\vec r)$ satisfies the following relation 
\be
\label{09}
\int \,\, {\rm d} \vec r \,\,{\tilde \nu}^{(0)\tilde\theta}(\vec r)\,\,{\tilde f}_{\tilde\theta\tilde\beta}(\vec r, \vec r^{\prime})= 0,
\ee 
which produces a set of equations that allows to determine $\Psi$. At this point, it is very important to notice that the formalism unveils the $U(1)$ hidden gauge symmetry of the physical model because the zero-mode does not generate a new constraint. Considering a general zero-mode,
\be
\label{009}
{\tilde\nu}^{(0)}=  \left(\begin{array}{ccccc}
a&b&c&d&-1
\end{array}\right),
\ee
we have the following set of differential equations
\ba
& &\int \,\, {\rm d}\vec r \,\,\left(b\delta(\vec r -\vec r^{\prime}) + \frac{\partial\psi}{\partial\rho} + d\alpha\delta(\vec r -\vec r^{\prime}) \right)
=0,\nonumber\\
& &\int \,\, {\rm d}\vec r \,\,\left(-a\delta(\vec r -\vec r^{\prime}) + \frac{\partial\psi}{\partial\theta} \right)=0,\\
& &\int \,\, {\rm d}\vec r \,\,\left(\frac{\partial\psi}{\partial\alpha} + d\rho\delta(\vec r -\vec r^{\prime})\right)=0,\nonumber\\
& &\int \,\, {\rm d}\vec r \,\,\left(\frac{\partial\psi}{\partial\beta} - a\alpha - c\rho\delta(\vec r -\vec r^{\prime})\right)=0.\nonumber
\ea
After a direct calculation, we obtain
\be
\Psi(\vec r) = -b\rho - \alpha d\rho +a\theta +\alpha a\beta +c\rho\beta.
\ee
In order to have a solution, we consider that $a=c=0$, then
\be
{\tilde\nu}^{(0)}=  \left(\begin{array}{ccccc}
0&1&0&1&-1
\end{array}\right),
\ee
and 
\be
\Psi(\vec r) = -(1+\alpha)\rho.
\ee

Now, we begin with the second step of the method to reformulate the model as a WZ gauge-invariant model. The zero-mode ${\tilde\nu}^{(0)}$ does not produce a constraint when contracted with the gradient of the symplectic potential, then,
\ba
\label{015}
& &\int {\rm d} \vec r^{\prime} \;\;\tilde\nu^{(0)\tilde\beta}(\vec r)\;\frac{\delta {\tilde V}^{(0)}(\vec r^{\prime})}{\delta {\tilde\xi}^{\tilde\beta}(\vec r)} = 0\,\,,\nonumber\\
& &\int {\rm d} \vec r^{\prime} \;\;\Big\{(1+\alpha) \rho (\partial_i^{\prime}\theta+\alpha\partial_i^{\prime}\beta) \partial_i^{\prime}\delta(\vec r^{\prime} - \vec r)  \nonumber\\
&+& \sum_{n=1}\left(\frac{\partial {\cal G}^{(n)}}{\partial\theta}+\frac{\partial {\cal G}^{(n)}}{\partial\beta}-\frac{\partial {\cal G}^{(n)}}{\partial\eta}\right)\Big\}=0\,\,. 
\ea
This expression produces a general differential equation, that allows the computation of all the correction terms as functions of $\eta$ enclosed into $G(\rho,\theta,\eta)$. To compute the first correction term as function of $\eta$, ${\cal G}^{(1)}$, we pick up the terms in equation (\ref{015}) with zeroth-order in $\eta$, thus
\ba
\label{16}
\int {\rm d} \vec r^{\prime} \;\;\left\{(1+\alpha)\rho(\partial_i^{\prime}\theta + \alpha \partial_i^{\prime}\beta)\partial_i^{\prime}\delta(\vec r^{\prime} - \vec r) -\frac{\partial {\cal G}^{(1)}}{\partial\eta}\right\}=0\,\,, \nonumber \\
\mbox{}
\ea
where $\partial_i^{\prime} =\frac{\partial}{\partial \vec r^{\prime}}$.
After a straightforward calculation, the linear correction term as function of $\eta$ is obtained as
\be
\label{17}
{\cal G}^{(1)}=(1+\alpha)\rho\left(\partial_i\theta+\alpha\partial_i\beta\right)\partial_i\eta.
\ee
For the quadratic correction terms in equation (\ref{015}), we have that
\be
\frac{\partial {\cal G}^{(1)}}{\partial\theta}+\frac{\partial {\cal G}^{(1)}}{\partial\beta}-\frac{\partial {\cal G}^{(2)}}{\partial\eta}=0.
\ee
After a direct calculation, the second-order correction term is obtained as
\be
\label{18}
{\cal G}^{(2)} =\frac{(1+\alpha)^2}{2}\rho\partial_i\eta\partial^i\eta.
\ee
For the cubic correction terms in equation (\ref{015}), we can write
\ba
\frac{\partial {\cal G}^{(2)}}{\partial\theta}+\frac{\partial {\cal G}^{(2)}}{\partial\beta}-\frac{\partial {\cal G}^{(3)}}{\partial\eta}&=& 0,\nonumber\\
\frac{\partial {\cal G}^{(3)}}{\partial\eta}&=& 0,
\ea
which allows to conclude that ${\cal G}^{(n)}=0$ for $n\geq3$.
Hence, the gauge-invariant first-order Lagrangian is written as
\ba
\label{26}
{\tilde{\cal L}}^{(0)} &=& -\rho(\dot \theta +\alpha\dot \beta)- (1+\alpha)\rho\dot\eta - {\tilde V}^{(0)},\nonumber\\
\mbox{}&=& -\rho(\dot \theta+ \dot \eta)- \alpha\rho(\dot\beta +\dot\eta) - {\tilde V}^{(0)}
\ea
where the symplectic potential is
\begin{widetext}
\ba
\label{27}
{\tilde V}^{(0)} &=&\frac{1}{2}(1-k)\rho (\partial_i \theta + \alpha\partial_i \beta) (\partial^i\theta + \alpha\partial^i\beta) + (1+\alpha)\rho\left(\partial_i\theta+\alpha\partial_i\beta\right)\partial_i\eta + \frac{(1+\alpha)^2}{2}\rho\partial_i\eta\partial^i\eta+ V(\rho),\nonumber\\
\mbox{}&=& \frac{1}{2}(1-k)\rho \left[\partial_i (\theta+\eta)
+  \alpha\partial_i (\beta+\eta)\right] \left[\partial^i (\theta+\eta) + \alpha\partial^i (\beta+\eta)\right]+ V(\rho)\,\,.
\ea
\end{widetext}
To complete the gauge-invariant reformulation of the model, we will compute the infinitesimal gauge transformation. In agreement with the method, the zero-mode ${\tilde \nu}^{(0)}$ is the generator of infinitesimal gauge transformations $(\delta{\cal O}=\varepsilon\tilde\nu^{(0)})$. Then,
\begin{eqnarray}
\label{30}
\delta \rho(\vec r,t) &=& 0,\nonumber\\
\delta \theta(\vec r,t) &=& \varepsilon(\vec r,t)\,\,\delta(\vec r - \vec r^{\prime}),\nonumber\\
\delta \alpha(\vec r,t) &=& 0,\\
\delta \beta(\vec r,t) &=& \varepsilon(\vec r,t)\,\,\delta(\vec r - \vec r^{\prime}),\nonumber\\
\delta \eta(\vec r,t)&=& -\varepsilon(\vec r,t)\,\,\delta(\vec r - \vec r^{\prime}),\nonumber
\end{eqnarray}
where $\varepsilon(\vec r,t)$ is an infinitesimal time-dependent parameter. In fact, under the infinitesimal transformations above, the invariant Hamiltonian $({\tilde V}^{(0)})$ changes as
\be
\label{31}
\delta{\tilde V}^{(0)}= 0.
\ee
Considering the following transformations, 
\ba
\label{TT}
\tilde\theta &=& \theta + \eta,\nonumber\\
\tilde\beta &=& \beta + \eta,
\ea
then the Lagrangian density, equation (\ref{26}), and the Hamiltonian, equation (\ref{27}), become
\ba
{\tilde{\cal L}}^{(0)} &=& -\rho(\dot{\tilde\theta}+\alpha\dot{\tilde\beta}) - {\tilde V}^{(0)}, \\
{\tilde V}^{(0)} &=& \frac{1}{2}\rho(\partial_i\tilde\theta+\alpha\partial_i\tilde\beta)(\partial^i\tilde\theta+\alpha\partial^i\tilde\beta)+ V(\rho)\,\,. \nonumber
\ea
These expressions are similar the original expressions for the Lagrangian in equation (\ref{002}), and the Hamiltonian in equation (\ref{03}), respectively. 

Thus, at this point, it is important to point out that exist a hidden symmetry into the rotational fluid model. Note that, in Section \ref{s2}, it was not possible to realize this symmetry. Another interesting feature discovered here is that for the set of the differential equations obtained we have no other solutions, {\it i.e.}, the model has, in fact, only one hidden symmetry. Based on the investigation done by some of us in Ref. \cite{01}, where the extra global symmetries proposed in \cite{02} are lifted to the local {\it status}, we can conclude that these extra global symmetries do not exist in the rotational fluid model.   Some other considerations will be depicted in the next section.


\section{Final Discussions}

From the Dirac point of view, a system classified as a gauge invariant theory is one that has only first-class constraints.  When a theory has second-class constraints, the gauge invariance can be recovered by converting the second-class constraints into first-class constraints.  In the literature there is a great variety of techniques, with pros and cons, to promote this kind of conversion.


In this work we are concerned not only with the gauge invariance of second-class systems but also in obtaining a theory dual equivalent to the original one.  We believe that dual embedding formalism is the most adequate technique because it is not affected by ambiguity problems related to the introduction of the WZ variables \cite{BN}.  Besides, as demonstrated by some of us, this method has the advantage that a convenient choice of a convenient zero-mode can lead to a theory dual equivalent to the parent one through the elimination of the WZ terms \cite{PD,ht}.  This means a new interpretation of the method.  However, as recently demonstrated by some of us \cite{jf} the choice of the zero-mode must obey some ``boundary conditions".   In other words, we can say that the physical coherence must guide us to choose the correct (or convenient) zero-mode.  But it is not an impediment to obtain, as mentioned just above, a whole family of dual equivalent actions.

Speaking in another way, we can say that this particular mapping between the parent action and the respective final gauge invariant theory can be interpreted as a kind of duality. The gauge invariance of the final actions obtained here was demonstrated via the Dirac analysis.  The characterization of the final action as a first-class system corroborates the success of this process of ``dualization".

Firstly in this paper we used a kind of toy model, the Proca model, to illustrate the procedure. Although being a toy model, the resulting ``dual" action is new.  After that, we apply the formalism to the non-linear sigma model and to the chiral Schwinger model.  In the NLSM, a hidden symmetry lying on the original phase space was disclosed, oppositely to other approaches \cite{BN,WN,BGB}, where the symmetry resides on the extended WZ phase space.

In the CSM, the chiral anomaly was eliminated and the gauge symmetry was recovered.
It is important to notice that this result was achieved introducing
one WZ field while other schemes thrive with the introduction of two or more WZ fields, which is the origin of the ambiguity problem.

Besides, we showed in the context of a non-Abelian model (the non-Abelian Proca model) that the dual embedding
formalism can be used without any restrictions with the algebra obeyed by the noninvariant model, while other constraint conversion
techniques work since the algebra was previously and necessarily taken into account.

We also have proposed a gauge-invariant version for the rotational fluid model. As a consequence, we have demonstrated that the hidden symmetry found is unique.  
Although we have studied the rotational fluid model with an extra term, which introduces dissipation into the model, the results are also valid without viscosity $(k=0)$.
We have noted that, although having dissipation, this fluid model presents hidden symmetry,
which does not belong to the other symmetries group obtained for the irrotational fluid model \cite{01}. Due to this, the local version of the extra global symmetries \cite{02} does not exist in the rotational fluid model, with dissipation or not. 
Furthermore, the physical meaning of the hidden symmetry can be interpreted. Consider a flow of fluid with viscosity  in a tube. It is well known that there are many layers \cite{symon}, each layer flowing with a specific velocity: the velocity of the layer at the center of the tube is maximum and, distant from the center, the velocity decreases in accordance with the increasing of the distance from the center. It also happens when vortices are present. The are many layers in a vortex and the one that is near to the center of the vortex has increased its velocity. Due to this, the velocity of each layer can be transformed into the velocity of the other layer by using the equation s in equation (\ref{TT}). Therefore, the dynamics of the fluid is preserved and governed by both Lagrangian and Hamiltonian which present the hidden symmetry.


\section{ Acknowledgments}

The authors would like to thank CNPq, FAPEMIG, FAPERJ and FAPESP, Brazilian Research Agencies, for financial support. EMCA would like to thank the kindness and hospitality of Departamento de F\' isica of Universidade Federal de Juiz de Fora, where part of this work has been accomplished.


\begin{thebibliography} {99}

\bibitem{PD} P. A. M. Dirac, Proc. Roy. Soc. A 257 (1960) 32;
P. A. M. Dirac, {\it Lectures on Quantum Mechanics}, Yeshiva University Press, New York, 1964;
A. Hanson, T. Regge and C. Teitelboim, {\it Constrained Hamiltonian Systems}, Academia Nazionale dei Lincei, Roma 1976;
K. Sundermeyer, {\it Constrained Dynamics, Lectures Notes in Physics}, Springer, New York, 1982, vol 169;
P. G. Bergmann, Phys. Rev. 75, (1949) 680; P. G. Bergmann, Phys. Rev. 89 (1953) 4.

\bibitem{ht}   M. Henneaux and C. Teiltelboim, {\it Quantization of Gauge Systems}, Princeton University Press, 1992.

\bibitem{FADDEEV}  L. D. Faddeev, Theor. Math. Phys. 1 (1970) 1.

\bibitem{FRADKIN}  E. S. Fradkin, G. A. Vilkovisky, Phys. Lett. B 55 (1975) 224; I. A. Batalin, G. A. Vilkovisky, Phys. Lett. B 69 (1977) 309.

\bibitem{BRST}  C. Becchi, A. Rouet and R. Stora, Ann. Phys. [N.Y.] 98 (1976) 287; Phys. Lett. B 52 (1974) 344;
I.V. Tyutin, Lebedev-preprint 39/1975.

\bibitem{order} N. Banerjee, B. Banerjee and S. Ghosh, Ann. Phys. 241 (1995) 237, and references therein.

\bibitem{RR} R. Rajaraman, Phys. Lett. B 154 (1985) 305.

\bibitem{BV} I. A. Batalin and G. A. Vilkovisky, Phys. Lett. B 102 (1981) 27; Phys. Lett. B 120 (1983) 166; Phys. Rev. D 28 (1983) 2587;
Phys. Rev. D 30 (1984) 508; Nucl. Phys. B 234, 106 (1984); J. Math. Phys. 26 (1985) 172.

\bibitem{FS}  L. Faddeev and S. L. Shatashivilli, Phys. Lett. B 167 (1986) 225.

\bibitem{BT} I. A. Batalin and E. S. Fradkin, Nucl. Phys. B 279 (1987) 514; Int. J. Mod. Phys. A  (1991) 3255.

\bibitem{IJMP} C. Wotzasek, Int. J. Mod. Phys. A 5 (1990) 1123.

\bibitem{many3}  M. Moshe and Y. Oz, Phys. Lett. B 224 (1989) 145; T. Fujiwara, Y. Igarashi and J. Kubo, Nucl. Phys. B 341 (1990) 695.

\bibitem{BN} J. Barcelos-Neto, Phys. Rev. D 55, (1997) 2265.

\bibitem{dualidade}  C. Montonen, D. I. Olive, Phys. Lett. B 72, 117 (1977); L. Alvarez-Gaume and F. Zamora, {\it Duality in quantum field theory and string theory}, hep-th/9709180; E. M. C. Abreu and M. Hott, Phys. Rev. D 62 (2000) 027702 and references therein.


\bibitem{ainrw} M. A. Anacleto, A. Ilha, J. R. S. Nascimento, R. F. Ribeiro and C. Wotzasek, Phys. Lett. B 504 (2001) 268.

\bibitem{iw}  A. Ilha and C. Wotzasek, Nucl. Phys. B 604 (2001) 426.

\bibitem{iw2}  A. Ilha and C. Wotzasek, Phys. Lett. B 519 (2001) 169.

\bibitem{iw3}  A. Ilha and C. Wotzasek, Phys. Lett.B 510 (2001) 329.

\bibitem{tpn} P. K. Townsend, K. Pilch and P. van Nieuwenhuizen, Phys. Lett. B 136 (1984) 38.

\bibitem{dj} S. Deser and R. Jackiw, Phys. Lett. B 139 (1984) 2366.

\bibitem{suecos}  S. E. Hjelmeland, U. Lindstr\"om, UIO-PHYS-97-03, May 1997, e-Print Archive: hep-th/9705122.

\bibitem{NLSM}   A. I. Vainshtein, V. I. Zakharov, V. A. Novikov and M. A. Shifman, Sov. J. Part. Nucl. 17 (1986) 204.

\bibitem{FJ} L. Faddeev and R. Jackiw, Phys. Rev. Lett. 60 (1988) 1692; N. M. J. Woodhouse, {\it Geometric Quantization}, Clarendon Press, Oxford, 1980.

\bibitem{BC} J. Barcelos-Neto and C. Wotzasek, Mod. Phys. Lett. A 7 (1992) 1172; Int. J. Mod. Phys. A 7 (1992) 4981.

\bibitem{amnot}   E. M. C. Abreu, A. C. R. Mendes, C. Neves, W. Oliveira and F. I. Takakura, Int. J. Mod. Phys. A  22 (2007) 3605.

\bibitem{gotay}  M. J. Gotay, J. M. Nester and G. Hinds, J. Math. Phys. 19 (11) (1978) 2388.

\bibitem{STRING}  J. Sonnenschein, Nucl. Phys. B 309 (1988) 752.

\bibitem{ANO} J. Ananias Neto, C. Neves and W. Oliveira, Phys. Rev.D 63 (2001) 085018; A. C. R. Mendes, C. Neves, W. Oliveira and D. C. Rodrigues, Nucl. Phys. B (Proc. Suppl.)  127 (2004) 170.

\bibitem{WN}  C. Neves and C. Wotzasek, J. Math. Phys. 34 (1993) 1807.

\bibitem{BGB}   N. Banerjee, S. Ghosh, and R. Banerjee, Nucl. Phys. B 417 (1994) 257.

\bibitem{JW1}  W. Oliveira and J. Ananias Neto, Int. J. Mod. Phys. A 12 (1997) 4895.

\bibitem{JW2} W. Oliveira and J. Ananias Neto, Nucl. Phys. B 533 (1998) 611.

\bibitem{HKP} S.-T. Hong, Y.-W. Kim and Y.-J. Park,  Phys. Rev. D 59 (1999) 114026; S.-T. Hong, Y.-W. Kim, Y.-J. Park, Mod. Phys. Lett. A 15 (2000) 55.

\bibitem{NW} C. Neves and C. Wotzasek, Phys. Rev. D 59 (1999) 125018.

\bibitem{NM} T. Maskawa and H. Nakajima, Prog. Theor. Phys. 56 (1976) 1295.

\bibitem{KR}  A. Kovner and B. Rosenstein, Phys. Rev. Lett. 59 (1987) 857.

\bibitem{JR}  R. Jackiw and R. Rajaraman, Phys. Rev. Lett. 54 (1985) 1219; {\it ibid} 54 (1985) 2060(E).

\bibitem{many}   R. Banerjee, Phys. Rev. Lett. 56 (1986) 1889; S. Miyake and K. Shizuya, Phys. Rev. D 36 (1987) 3781; K. Harada and I. Tsutsui, Phys. Lett. B 171 (1987) 311.

\bibitem{RR1}  R. Rajarama, Phys. Lett. B 162 (1985) 148.

\bibitem{LR}  J. Lott and R. Rajarama, Phys. lett. B 165 (1985) 321.

\bibitem{01} A. C. R. Mendes, C. Neves, W. Oliveira and F. I. Takakura, J. Phys. A: Math. Gen.  37 (2004) 1927.

\bibitem{02} D. Bazeia, R. Jackiw, Annals Phys. 270 (1998) 246; {\it Field
dependent diffeomorphism symmetry in diverse dynamical systems}, e-Print Archive:
hep-th/9803165; D. Bazeia, Phys. Rev. D 59 (1999) 085007.



\bibitem{RJ} R. Jackiw, V.P. Nair, S.Y. Pi, A. P. Polychronakos. MIT-CTP-3509, J. Phys. A 37 (2004) R327; e-Print: hep-ph/0407101.

\bibitem{Lin} C. C. Lin, Int. School of physics E. Fermi XXI, G. Careri, ed. (Academic Press, New York, 1963).

\bibitem{NW2} C. Neves, W. Oliveira, Phys. Lett. A 321 (2004) 267; e-Print: hep-th/0310064.

\bibitem{symon} K. R. Symon, Mechanics, third edition, Ed. Addison-Wesley publishing company, 1971(chapter 8).

\bibitem{jf}  E. M. C. Abreu, A. C. R. Mendes, C. Neves, W. Oliveira C. Wotzasek and L.M.V. Xavier ``Duality considerations about the Maxwell-Podolsky theory through the symplectic embedding formalism and spectrum analysis", arXiv: 0812.0950 [hep-th], to appear in
Int. J. Mod. Phys. A.





\end{thebibliography}
\end{document}